\documentclass[reprint,
aps,superscriptaddress,twocolumn,showpacs,longbibliography]{revtex4-2}
\usepackage{graphicx,color}
\usepackage{amsmath,amsfonts,enumerate,amsthm,amssymb,bbm}
\usepackage[colorlinks=true,citecolor=blue,linkcolor=magenta]{hyperref}
\usepackage{multirow}
\usepackage{bbold}
\usepackage{braket}
\usepackage{soul}
\usepackage[caption = false]{subfig}
\usepackage{scrextend}
\usepackage{xcolor}
\usepackage{dsfont}
\usepackage{nicefrac}
\usepackage[linesnumbered,ruled,vlined]{algorithm2e}
\usepackage{quantikz}
\usepackage{apptools}

\mathchardef\ordinarycolon\mathcode`\:
\mathcode`\:=\string"8000
\begingroup \catcode`\:=\active
  \gdef:{\mathrel{\mathop\ordinarycolon}}
\endgroup

\theoremstyle{plain}
\newtheorem{thm}{Theorem}
\newtheorem{corol}{Corollary}
\newtheorem{lem}{Lemma}

\theoremstyle{definition}

\theoremstyle{remark}
\newtheorem{rmk}{Remark}

\AtAppendix{\counterwithin{thm}{section}}
\AtAppendix{\counterwithin{corol}{section}}
\AtAppendix{\counterwithin{lem}{section}}
\AtAppendix{\counterwithin{propos}{section}}
\AtAppendix{\counterwithin{defn}{section}}
\AtAppendix{\counterwithin{rmk}{section}}
\AtAppendix{\counterwithin{ex}{section}}

\def\<{\langle}

\def\>{\rangle}
\def\<{\langle}
\DeclareMathOperator{\Tr}{Tr}

\newcommand{\thv}{\vec{\theta}}

\newcommand{\alv}{\vec{\alpha}}
\newcommand{\gamv}{\vec{\gamma}}

\newcommand{\norm}[1]{\left\lVert#1\right\rVert}

\renewcommand{\ket}[1]{|#1\rangle}               
\renewcommand{\bra}[1]{\langle #1|}              

\renewcommand{\vec}[1]{\boldsymbol{#1}}  

\begin{document}

\title{Dynamical simulation via quantum machine learning with provable generalization}

\author{Joe Gibbs}
\thanks{The first two authors contributed equally to this work.}
\affiliation{AWE, Reading, UK.}

\author{Zo\"{e} Holmes}
\thanks{The first two authors contributed equally to this work.}
\affiliation{Information Sciences, Los Alamos National Laboratory, Los Alamos, NM, USA.}

\author{Matthias C.~Caro}
\affiliation{Department of Mathematics, Technical University of Munich, Garching, Germany.}
\affiliation{Munich Center for Quantum Science and Technology (MCQST), Munich, Germany.}
\affiliation{Dahlem Center for Complex Quantum Systems, Freie Universität Berlin, Berlin, Germany.}

\author{Nicholas Ezzell}
\affiliation{Information Sciences, Los Alamos National Laboratory, Los Alamos, NM, USA.}
\affiliation{Department of Physics \& Astronomy, University of Southern California, Los Angeles, CA, USA}

\author{Hsin-Yuan Huang}
\affiliation{Institute for Quantum Information and Matter, Caltech, Pasadena, CA, USA.}
\affiliation{Department of Computing and Mathematical Sciences, Caltech, Pasadena, CA, USA}

\author{Lukasz Cincio}
\affiliation{Theoretical Division, Los Alamos National Laboratory, Los Alamos, NM, USA.}

\author{Andrew T. Sornborger} 
\affiliation{Information Sciences, Los Alamos National Laboratory, Los Alamos, NM, USA.}

\author{Patrick~J.~Coles} 
\affiliation{Theoretical Division, Los Alamos National Laboratory, Los Alamos, NM, USA.}

\date{\today}

\begin{abstract}
Much attention has been paid to dynamical simulation and quantum machine learning (QML) independently as applications for quantum advantage, while the possibility of using QML to enhance dynamical simulations has not been thoroughly investigated. Here we develop a framework for using QML methods to simulate quantum dynamics on near-term quantum hardware. We use generalization bounds, which bound the error a machine learning model makes on unseen data, to rigorously analyze the training data requirements of an algorithm within this framework. This provides a guarantee that our algorithm is resource-efficient, both in terms of qubit and data requirements. Our numerics exhibit efficient scaling with problem size, and we simulate 20 times longer than Trotterization on IBMQ-Bogota.
\end{abstract}

\maketitle

\paragraph*{Introduction}

The exponential speedup of dynamical quantum simulation provided the original motivation for quantum computers~\cite{feynman1982simulating,lloyd1996universal}. In the long term, large-scale quantum simulations are expected to transform fields such as materials science, chemistry, and high-energy physics. Nearer term, since efficient classical dynamical simulation methods are lacking (in contrast to those for computing static quantum properties like electronic structure), dynamical simulation may plausibly be one of the first applications to see quantum advantage.

Achieving near-term quantum advantage for dynamics will require long-time simulations on Noisy Intermediate-Scale Quantum (NISQ) hardware~\cite{preskill2018quantum}. Standard methods like Trotterization grow the circuit depth in proportion to the simulation time, ultimately running into the decoherence time of the NISQ device~\cite{berry2015simulating,low2019hamiltonian}. Fast-forwarding methods for long-time simulations on NISQ devices have recently been introduced~\cite{cirstoiu2020variational,commeau2020variational,gibbs2021long, geller2021experimental}, but are limited by various inefficiencies (e.g., qubit and data requirements). 
Here, we address these inefficiencies, potentially opening the door for near-term quantum advantage.

In parallel to these developments, quantum machine learning (QML)~\cite{biamonte2017quantum, schuld2021quantum} has emerged as another potential application for quantum advantage~\cite{huang2021quantumadvantage}. At its core, QML involves using classical or quantum data to train a parameterized quantum circuit.  
A number of promising paradigms for training are being pursued, including variational quantum algorithms using training data~\cite{cerezo2020variationalreview}, quantum generative adversarial networks~\cite{lloyd2018quantum, zoufal2019quantum} and quantum kernel methods~\cite{schuld2021quantum} (to name just a few). 
Here we seek to combine the potential of QML and dynamical simulation, by leveraging recent advances in QML to reduce resource requirements for dynamical simulation.

To assess the scalability of QML methods, as well as their applicability to real world problems, it is critical to understand their training data requirements, quantified by so-called generalization bounds~\cite{caro2020pseudo, bu2021onthestatistical, gyurik2021structural, abbas2020power, du2021efficient, caro2021encodingdependent, chen2021expressibility, popescu2021learning, caro2021generalization, cai2022sample, caro2022outofdistribution, poland2020no, sharma2020reformulation, Volkoff2021Universal}. These provide bounds on the error a machine learning model makes on unseen data, as a function of the amount of data the model is trained on and of the training performance. In this letter, we assess the training data requirements of QML approaches to dynamical simulation.

Our analysis provides the groundwork for a new QML-inspired algorithm for dynamical simulation that we call the `Resource-Efficient Fast Forwarding' algorithm (REFF). This algorithm uses training data to learn a circuit that allows for fast-forwarding, where long-time simulations can be performed using a fixed-depth circuit. The REFF algorithm is efficient in the amount of training data required. It is also qubit-efficient in the sense that simulating an $n$ qubit system only requires $n$ qubits, in contrast to earlier work~\cite{cirstoiu2020variational} which required $2n$ qubits. We use generalization bounds to rigorously lower bound the final simulation fidelity as a function of the amount of training data used, the optimization quality, and the simulation time. This analysis is complemented by numerical implementations, as well as a demonstration of our algorithm on IBMQ-Bogota.

\medskip

\paragraph*{General Framework} 

\begin{figure}[t]
\centering
\includegraphics[width =\columnwidth]{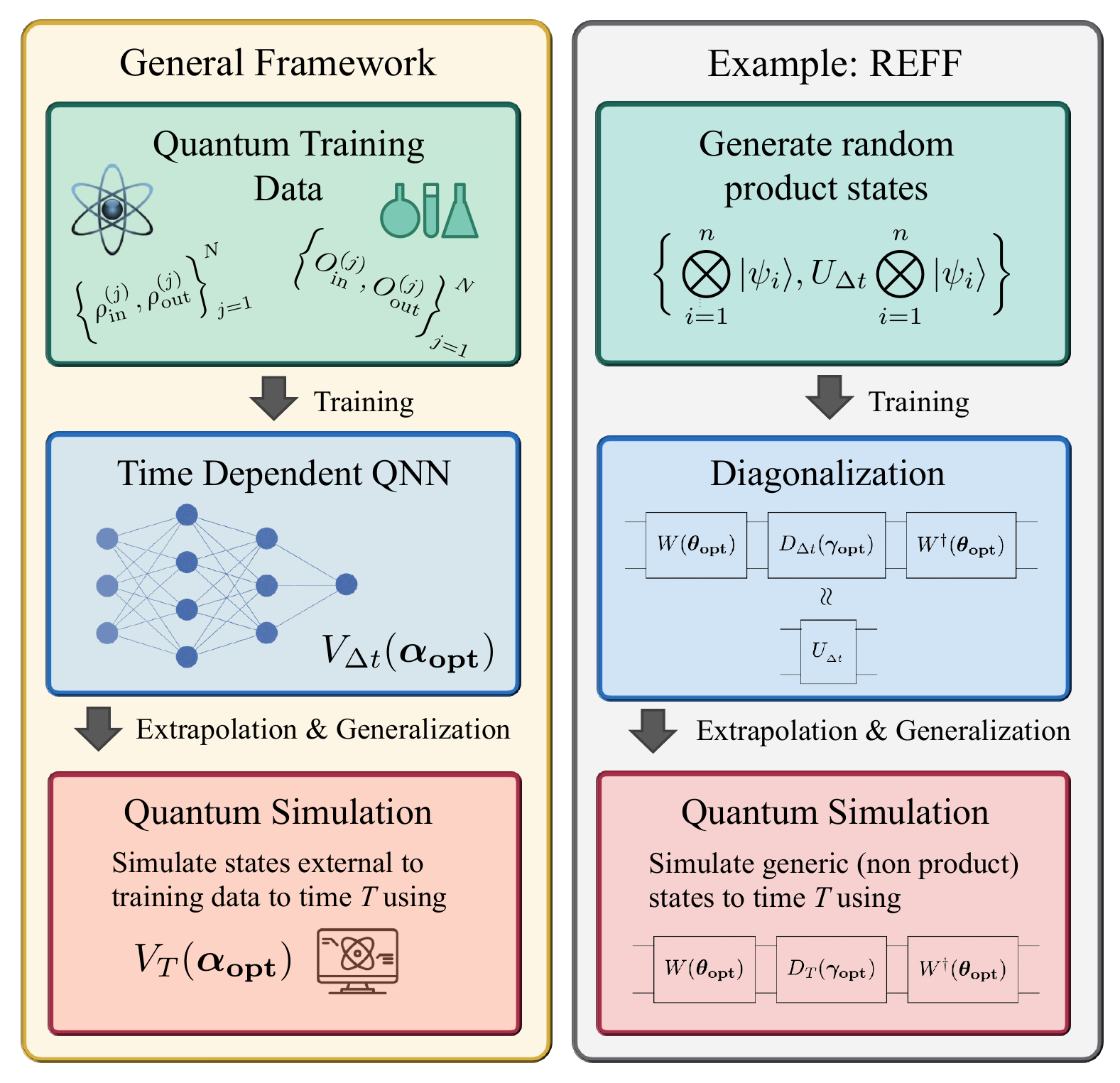}
\vspace{-6mm}
\caption{\small \textbf{QML framework for dynamical simulation.} Our general framework (left panel) consists of using quantum training data, e.g., composed of input-output state pairs and/or input-output observable pairs, to train a time-dependent QNN, $V_{t}(\alv)$. Typically the training occurs at a short time $\Delta t$, resulting in the trained QNN, $V_{\Delta t}(\alv_{\rm opt})$. The evolution of the system at some longer time $T$ is extrapolated via $V_{T}(\alv_{\rm opt})$. The Resource Efficient Fast Forwarding (REFF) algorithm (right panel) is an illustrative example of this framework. The training data consists of Haar-random product states as inputs and then the time-evolved states as outputs, i.e., evolved under $U_{\Delta t} \approx \exp(- i H \Delta t)$. The time-dependent QNN is a parameterized quantum circuit in a diagonal form: $W(\vec{\theta}) D_t(\vec{\gamma}) W^\dagger(\vec{\theta})$. Hence training the QNN amounts to approximately diagonalizing the short-time evolution $U_{\Delta t}$, resulting in the trained QNN: $W(\vec{\theta}_{\rm opt}) D_{\Delta t}(\vec{\gamma}_{\rm opt}) W^\dagger(\vec{\theta}_{\rm opt})$. This model can simulate the evolution of arbitrary input states up to time $T$ via $W(\vec{\theta}_{\rm opt}) D_T(\vec{\gamma}_{\rm opt}) W^\dagger(\vec{\theta}_{\rm opt})$.}
\label{fig:Schematic}
\end{figure}

Given a set of initial states $\mathcal{S}_{\rho}$ and an $n$-qubit Hamiltonian $H$, the goal of dynamical simulation is to predict the evolution of a set of observables $\mathcal{S}_O$ up to time $T$. A promising approach in the NISQ era is to use QML to fit a time-dependent quantum model that can be extrapolated to long simulation times using a short-depth quantum circuit~\cite{cirstoiu2020variational,kokcu2021fixed,commeau2020variational,gibbs2021long}. This is valuable in the NISQ era since high noise levels constrain the depth of circuits that may be simulated. More concretely, the aim is to find some time-dependent Quantum Neural Network (QNN), $V_{t}(\vec{\alpha})$, and optimized parameters, $\alv_{\rm opt}$, such that for any time $t < T$, 
\begin{equation}\label{eq:aim}
    \Tr[ V_{t}(\alv_{\rm opt}) \rho V_{t}(\alv_{\rm opt})^\dagger O ] \approx \Tr[ e^{-i H t} \rho e^{i H t}  O ]
\end{equation}
for any $\rho \in \mathcal{S}_{\rho}$ and $O \in \mathcal{S}_{O}$. 
Possible time-dependent QNNs include those formed from the Newton-Cartan decomposition of $H$~\cite{kokcu2021fixed, steckmann2021simulating}, or a diagonalization of $H$~\cite{commeau2020variational, radha2021quantum, horowitz2022quantum}, or of the propagator for short-time evolution~\cite{cirstoiu2020variational, gibbs2021long}. Fig.~\ref{fig:Schematic} depicts this framework.

In training $V_{t}(\vec{\alpha})$, the most appropriate choice in training data will depend on the set of initial states $\mathcal{S}_\rho$ and observables $\mathcal{S}_O$, but typically will be generated by the properties of the system at short times. For example, if one is interested in only knowing the evolution of a single observable, the training data could consist of the evolution of this observable for some subset of the target states up to some short time $\Delta t$. Alternatively, if one is interested in simulating the evolution of any possible $n$ qubit observable, it would be natural to use training data consisting of $N$ pairs of input-output states $\mathcal{D}(N) = \{(\ket{\Psi^{(j)}},\ket{\Phi^{(j)}}) \}_{j=1}^{N}$ where the output states are generated by evolving the input states for some short time, i.e., $\ket{\Phi^{(j)}}=U_{\Delta t}\ket{\Psi^{(j)}}$ where $U_{\Delta t}\approx e^{-i H \Delta t}$ is a gate sequence (such as a Trotterization) that approximates the true time evolution.

The training data is initially used to train the QNN to reproduce the properties of the system at short times, i.e., to ensure that Eq.~\eqref{eq:aim} holds for times $t \leq \Delta t$. Crucially, while trained on a subset of the target states and/or observables, the hope is that the learned QNN \textit{generalizes} to the unseen target data, i.e., it well reproduces Eq.~\eqref{eq:aim} for short times ($t \lesssim \Delta t$) for any $\rho \in \mathcal{S}_{\rho}$ and $O \in \mathcal{S}_{O}$.
The properties of the system at some longer time $T$ can then be extrapolated via $V_{T}(\alv_{\rm opt})$. 

Generalization bounds quantify the performance of a QNN on unseen data, after optimization on a limited training set. In particular, the generalization error measures how much the performance on new data differs from the performance on the training data. 
Ref.~\cite{caro2021generalization} has shown that if the parameterized quantum circuit has $K$ trainable local gates, the generalization error scales at worst as $\tilde{\mathcal{O}}(\sqrt{K/N})$. Crucially, as established in Ref.~\cite{caro2022outofdistribution}, these training states may be product states. 
Below we use these quantum generalization bounds to quantify the performance of dynamical simulation via QML. 

\medskip

\paragraph*{REFF algorithm} 

For the remainder of this letter, we focus on a time-dependent QNN of the following form: 
\begin{align}\label{eq:VHDansatz}
V_{ t}(\alv)= W(\thv)D_t(\gamv)W^\dagger(\thv) \,,
\end{align}
where $\alv = (\thv, \gamv)$ and $W( \thv)$ is a time-independent unitary. $D_t(\gamv)$ is a time-dependent unitary that is diagonal in the standard basis. Since $D$ is diagonal, $M$ applications of $D_{t}$ are equivalent to one application of $D_{Mt}$, i.e., $D_{t}(\gamv)^M = D_{M t}(\gamv)$ for any positive integer $M$.
We will train $V_{ t}(\alv)$ at time $t=\Delta t$ to obtain the trained QNN, $V_{\Delta t}(\alv_{\rm opt})$.
If this well-approximates the target unitary, i.e., $V_{\Delta t}(\alv_{\rm opt}) \approx U_{\Delta t}$, then we have learned an approximate diagonalization of $U_{\Delta t}$ and the long-time simulation $e^{-iH T}$ can be approximated using the fixed depth circuit $V_T(\alv_{\rm opt})$~\cite{cirstoiu2020variational,gibbs2021long}.
Below we formally bound the fidelity between the simulated evolution and the exact evolution.

In what follows we consider the (most difficult) task of learning the dynamics for all input states and all observables, i.e., where $S_{\rho}$ is the whole Hilbert space and $S_O$ is the set of all positive operator valued measure (POVM) elements. 
(We remark, however, that our analysis can be extended to dynamics within subspaces, such as subspaces that preserve certain symmetries.) 
For this task, as discussed above, a natural choice in training data would be $N$ pairs of input-output states $\mathcal{D}(N) = \{(\ket{\Psi^{(j)}},U_{\Delta t}\ket{\Psi^{(j)}}) \}_{j=1}^{N}$. A simple choice for the input training states $\ket{\Psi^{(j)}}$ would be to use (Haar) random $n$-qubit states. However, such random states will typically be highly entangled, requiring deep circuits to prepare, and thus are unsuitable for NISQ hardware. A more promising approach is to use tensor products of (Haar) random single-qubit states, which are preparable using only single-qubit gates, and so induce less noise. Thus we suppose that $U_{\Delta t}$ is learned using a set of \textit{product} input states, i.e., that the training data has the form
\begin{equation}
    \mathcal{D}_{\rm P}(N) :=  \{(\ket{\Psi_{\rm P}^{(j)}},\ket{\Phi_{\rm P}^{(j)}}) \}_{j=1}^{N} \ \ \text{with}  \ \ \ket{\Psi_{\rm P}^{(j)}}  = \bigotimes_{i=1}^n \ket{\psi_i^{(j)}}\, ,
\end{equation}
where the states $\{ \ket{\psi_i^{(j)}} \}_{i=1}^{N}$ are independently drawn from the single-qubit Haar distribution~\footnote{We can replace the assumption of independent Haar-random single-qubit states by an assumption of independent states drawn at random according to a $1$-qubit $2$-design and our error analysis will still hold.}.

There is freedom in how exactly this training data is used to learn $U_{\Delta t}$ but a natural approach would be to minimize the distance between the target output state $\ket{\Phi_{\rm P}^{(j)}} = U_{\Delta t} \ket{\Psi_{\rm P}^{(j)}}$ and the hypothesized output state $V_{\Delta t}(\alv) \ket{\Psi_{\rm P}^{(j)}}$, averaged over the $N$ states in the training set. That is, we minimize the following cost 
\begin{equation}
    C_{\mathcal{D}_{\rm P}(N)}^{\rm G}( \vec{\alpha}) 
    = \frac{1}{4N}\sum_{j = 1}^N \big\lVert \ket{\Phi_{\rm P}^{(j)}}\bra{\Phi_{\rm P}^{(j)}}  - V \ket{\Psi_{\rm P}^{(j)}}\bra{\Psi_{\rm P}^{(j)}} V^\dagger \big\rVert_1^2 \, ,
\end{equation}
where for compactness we write $V_{\Delta t}(\vec{\alpha}) \equiv V$ and $\lVert\cdot\rVert_1$ denotes the trace norm.
We can rewrite this cost in terms of the fidelity as
\begin{align}
C_{\mathcal{D}_P(N)}^{\rm G}( \vec{\alpha})    &= \frac{1}{N}\sum_{j = 1}^N \left( 1 - \left\lvert \bra{\Phi_{\rm P}^{(j)}} V \ket{\Psi_{\rm P}^{(j)}} \right\rvert^2 \right) \, ,
\label{eq:GlobalCost}
\end{align}
which can be measured on a quantum computer using the Loschmidt echo circuit~\cite{sharma2020reformulation} shown in Appendix~\ref{Ap:Numerics}.

While natural and intuitive, Eq.~\eqref{eq:GlobalCost} is a \text{global} cost~\cite{cerezo2020cost} since it is measured via the global measurements $\mathds{1} - \ket{\Psi_{\rm P}^{(j)}}\bra{\Psi_{\rm P}^{(j)}}$ on all $n$ qubits of the states $V^\dagger\ket{\Phi_{\rm P}^{(j)}}\bra{\Phi_{\rm P}^{(j)}} V$ for $j=1,\ldots,N$. 
Hence, it encounters exponentially vanishing gradients, known as barren plateaus~\cite{mcclean2018barren,cerezo2020cost,cerezo2020impact,arrasmith2020effect,Holmes2020Barren,holmes2021connecting,volkoff2021large,sharma2020trainability,pesah2020absence,uvarov2020barren,marrero2020entanglement,patti2020entanglement}. To mitigate such trainability issues, we advocate instead training using a local version of the cost of the form 
\begin{equation}\label{eq:localcost}
    C_{\mathcal{D}_{\rm P}(N)}^{\rm L}( \vec{\alpha})  
    = \frac{1}{N}\sum_{j = 1}^N \Tr\left[  V^\dagger \ket{\Phi_{\rm P}^{(j)}}\bra{\Phi_{\rm P}^{(j)}} V O_L^{(j)} \right] \, ,
\end{equation}
with $O_L^{(j)}
     := \mathds{1} - \frac{1}{n}\sum_{i=1}^n \ket{\psi_i^{(j)}}\bra{\psi_i^{(j)}} \otimes \mathds{1}_{\overline{i}}$, where $\overline{i}$ denotes the set of all qubits except for $i$. This cost is faithful~\cite{khatri2019quantum}, i.e., vanishing if and only if $U_{\Delta t} = V$, but crucially is also trainable as long as the ansatz is not too deep~\cite{cerezo2020cost}.

We call this algorithm, which uses the local product state cost, Eq.~\eqref{eq:localcost}, to learn a diagonalization QNN of the short time evolution unitary of a system and thereby fast-forward its evolution, the Resource-Efficient Fast Forwarding (REFF) algorithm. The algorithm is both efficient in terms of qubit usage (requiring only $n$ qubits to simulate an $n$-qubit system) and, as shown below, in terms of quantum data usage.

\medskip

\paragraph*{Simulation Error Bounds}\label{sec:learning-errors}

An operationally meaningful measure of the quality of the simulation via $V_T(\alv_{\rm opt})$ is given by the average simulation fidelity~\footnote{Note that by relating fidelity to trace distance and then invoking the operational meaning of the latter in terms of observable differences~\cite{nielsen2000quantum}, the simulation fidelity gives rise to an upper bound on the difference between the left- and right-hand sides in Eq.~\eqref{eq:aim}.}
\begin{equation}\label{eq:AverageFidelity}
    \overline{F}(\alv_{\rm opt}, T) =  \int_\psi | \bra{\psi}V_T(\alv_{\rm opt})^\dagger  e^{-iHT} \ket{\psi} |^2 ~\mathrm{d}\psi \, , 
\end{equation}
where the integral is over states $\ket{\psi}$ chosen according to the $n$-qubit Haar measure.
In this section, we lower bound the final simulation fidelity $\overline{F}(\alv_{\rm opt}, T)$ for the REFF algorithm, allowing for an arbitrary optimization procedure. Our bound depends on the time simulated, $T$, the amount of training data, $N$, the learning error over the training data (i.e., the minimum cost achieved, $C_{\mathcal{D}_{\rm P}(N)}^{\rm G}(\alv_{\rm opt})$ or $C_{\mathcal{D}_{\rm P}(N)}^{\rm L}( \alv_{\rm opt})$), and the error incurred from approximating the short time evolution $e^{- i H \Delta t}$ with the gate sequence $U_{\Delta t}$, that is $ \epsilon = \lVert U_{\Delta t} - e^{- i H \Delta t}\rVert_2$.

\begin{thm}[Simulation error for product-state training -- Informal]\label{thm:product-training-global-cost}
    Consider a QNN $V_{t}(\alv)$ given by Eq.~\eqref{eq:VHDansatz} and composed of $K$ parameterized local gates. When trained with the global cost $C_{\mathcal{D}_{\rm P}(N)}^{\rm G}$ using training data $\mathcal{D}_{\rm P}(N)$, the simulation fidelity after time $T = M\Delta t$, for a positive integer $M$, satisfies
    \begin{align}\label{eq:FidBoundGlobal}
        \overline{F}(\alv_{\rm opt}, T)
        &\geq 1 - 8 M^2 \left[\tilde{\epsilon}^2 + C_{\mathcal{D}_{\rm P}(N)}^{\rm G}(\alv_{\rm opt})\right] \nonumber \\
        &\hphantom{\geq 1~}- \mathcal{O} \left(M^2 f(K, N) \right) \, ,
    \end{align}
    with high probability over the choice of random product state data.
    Here $f(K, N) := \sqrt{\frac{K\log\left(K\right)}{N}}$ and $\tilde{\epsilon} := \frac{\epsilon}{2\sqrt{(2^n +1})}$.
    
    Alternatively, if the local cost $C_{\mathcal{D}_{\rm P}(N)}^{\rm L}$ is used for training, Eq.~\eqref{eq:FidBoundGlobal} holds with $C_{\mathcal{D}_{\rm P}(N)}^{\rm G}(\alv_{\rm opt}) \rightarrow n C_{\mathcal{D}_{\rm P}(N)}^{L}(\alv_{\rm opt})$ and $\mathcal{O} \left( M^2 f(K, N) \right)  \rightarrow \mathcal{O} \left(n M^2 f(K, N) \right)$.
\end{thm}

Theorem~\ref{thm:product-training-global-cost} implies that the fast-forwarded simulation fidelity deviates from $1$ at worst quadratically in the number of fast forwarding steps, $M$. Moreover, inverting Eq.~\eqref{eq:FidBoundGlobal} provides a means of bounding the number of product state training pairs and the minimal cost function sufficient to guarantee a given desired simulation fidelity and total simulation time.

In particular, Theorem~\ref{thm:product-training-global-cost} implies that a high fidelity may be achieved whenever the number of training pairs $N$ is effectively of the same order as $M^4 Kn^2$, i.e., scales polynomially in the product of the number of fast forwarding steps, the number of parameters used for the diagonalization, and the system size. 
Thus, the success of this QML-inspired approach to simulation depends critically on the number of parameters required to approximately diagonalize the short-time evolution of a system. 

For example, Ref.~\cite{verstraete2009quantum} analytically constructs the circuits required to exactly diagonalize the Ising chain. These circuits, which can simulate the XY model and Kitaev’s honey–comb lattice, require $n$ gates for the diagonal matrix $D$ and $\mathcal{O}(n^2)$ gates for $W$. Thus, assuming the correct discrete structure is known (or can be approximately found variationally), such models are captured by ans\"{a}tze with only a polynomial number of parameters. This provides hope that other systems may similarly be diagonalized with favorable parameter number scalings. 

\medskip
\paragraph*{Implementations}

In practice, even less training data than that suggested by our theoretical bounds may suffice. Here we numerically investigate the minimal training data required for two oft-studied Hamiltonians. We train on random product input states, and to simplify notation we write $C^{\rm L/G}_{\rm REFF} \equiv C^{\rm L/G}_{\mathcal{D}_{\rm P}(N)}( \vec{\alpha})$. To evaluate the quality of the learned diagonalization and the resulting simulation, we use the average fidelity $\overline{\mathcal{F}}_M$ between the learned diagonalization and a second order Trotterized unitary $U_{\Delta t}^M$. (This is similar to Eq.~\eqref{eq:AverageFidelity}, but with $U_{\Delta t}^M$ replacing $e^{-iH T}$.)

\begin{figure}[t]
\centering
\includegraphics[width =\columnwidth]{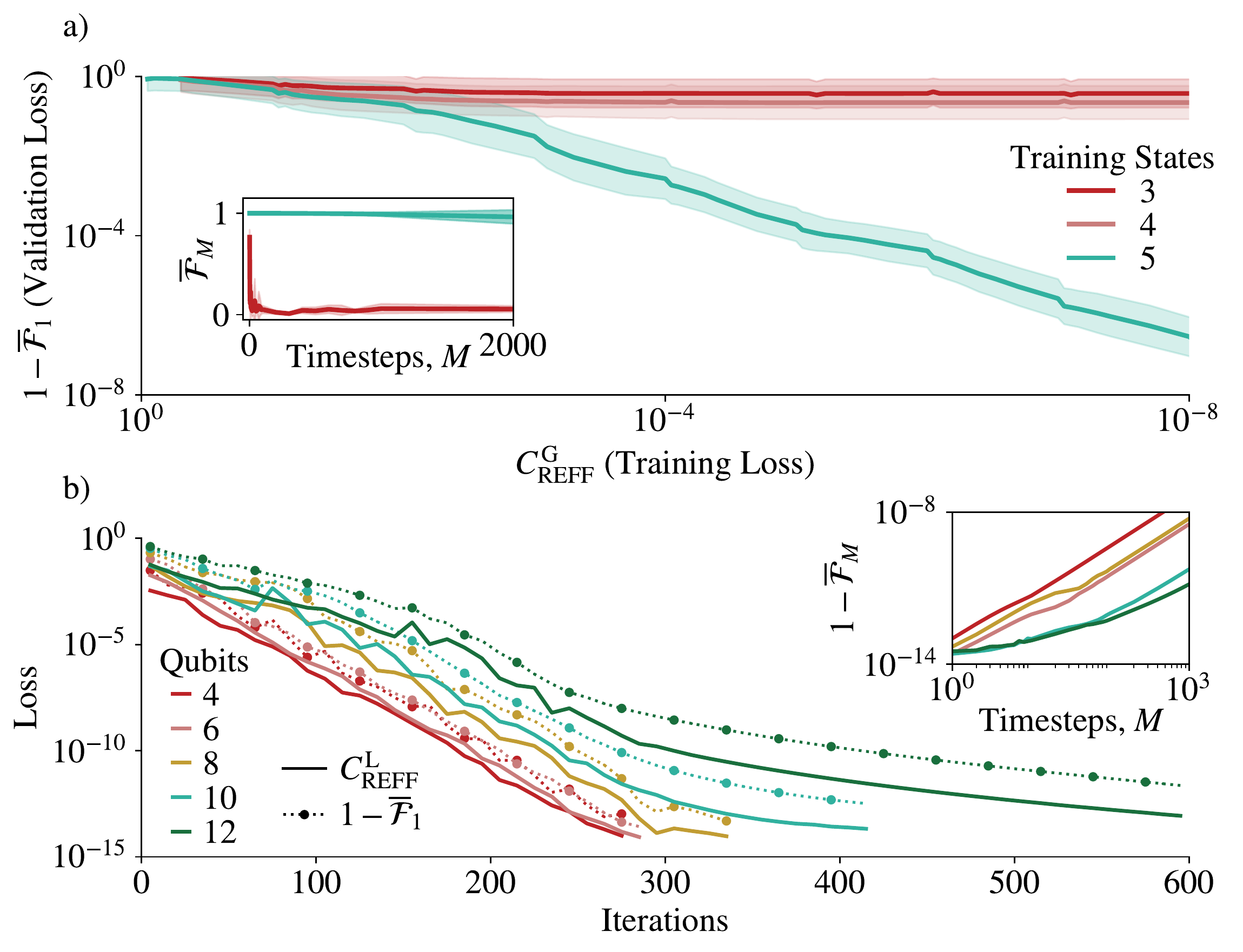}
\vspace{-6mm}
\caption{\small \textbf{Numerical simulations.} a) REFF is used to diagonalize the 4-qubit Heisenberg Hamiltonian with periodic boundary conditions. Only 5 Haar-random product training states are required to generalize over the whole Hilbert space, measured by the decrease in simulation infidelity $1-\overline{\mathcal{F}}_1$. The inset shows that the average fidelity $\overline{\mathcal{F}}_M \equiv \overline{\mathcal{F}} (\alv_{\rm opt}, M \Delta t)$ in this case remains over 0.95 for 2000 time steps. b) REFF is applied to increasing sizes of the XY model. For all sizes tested, a single Haar-random product training state was sufficient to achieve a machine precision simulation infidelity. For each system size, the ansatz is saved upon reaching $C^{\rm L}_{\rm REFF} = 10^{-14}$ and used to fast-forward the evolution, as shown in the inset.}
\label{fig:main_numerics}
\end{figure}

Let us consider the 4-qubit Heisenberg Hamiltonian $H = \sum_{i = 1}^4 \vec{S}_i \cdot \vec{S}_{i+1}$ with periodic boundary conditions.
For a set of input states composed of 3, 4, or 5 Haar-random product states, we minimized the $C^{\rm G}_{\rm{REFF}}$ cost and simultaneously tracked the average gate fidelity between the ansatz~\footnote{For more details on the ansatz we used see Appendix~\ref{Ap:Numerics} of the Supplementary Material.} and the Trotter unitary (with $\Delta t = 0.1$), as shown in Fig.~\ref{fig:main_numerics}a). The optimization was repeated for 10 different sets of training data, from which we computed the geometric mean and standard deviation (arithmetic mean and standard deviation in logspace). We found that 5 input states were sufficient to perform a full Hilbert space diagonalization, with all 10 runs of 5 training states achieving $1-\overline{\mathcal{F}}_1 < 10^{-6}$ once $C^{\rm G}_{\rm{REFF}}$ had reached $10^{-8}$. As shown in the inset, the QNN trained with $k = 5$ states generalized well, producing a long-time high fidelity simulation of the Hamiltonian evolution.

Next, we study the scaling with system size of the number of training states required to simulate the XY model. Fig.~\ref{fig:main_numerics}b) shows the results of using REFF to diagonalize and fast-forward the XY Hamiltonian, $H=\sum_{i=1}^{n-1} X_i X_{i+1} + Y_i Y_{i+1}$ with open boundary conditions, for $n \in \{4,6,\ldots, 12\}$. Here we use the local training cost, Eq.~\eqref{eq:localcost}, to help mitigate the exponential suppression of gradients as the system size grows due to Barren Plateaus. In all system sizes tested, as shown by the overlap of the lines and markers, only a single training state was required to diagonalize over the entire Hilbert space.
The inset of Fig.~\ref{fig:main_numerics}b) indicates that the final simulation error, as measured by $1-\overline{\mathcal{F}}_M$, scales sub quadratically with time, as predicted. In addition, we observe efficient (i.e., polynomial) scaling with $n$ for the gate count required for diagonalization. For further elaboration on this point, and as well as additional implementation details, see Appendix~\ref{Ap:Numerics} of the Supplementary Material. 
We note that our simulation fidelities are computed with respect to the Trotterized unitary, to quantify the quality of the learning of the target unitary. The Trotterization has an associated Trotter error that will cause a deviation from the true dynamics of the Hamiltonian, however increasing the order of the Trotter approximation can arbitrarily decrease this error. This effect on the simulation fidelity is numerically explored further in Appendix~\ref{Ap:Numerics}.

\begin{figure}[t!]
\centering
\includegraphics[width =\columnwidth]{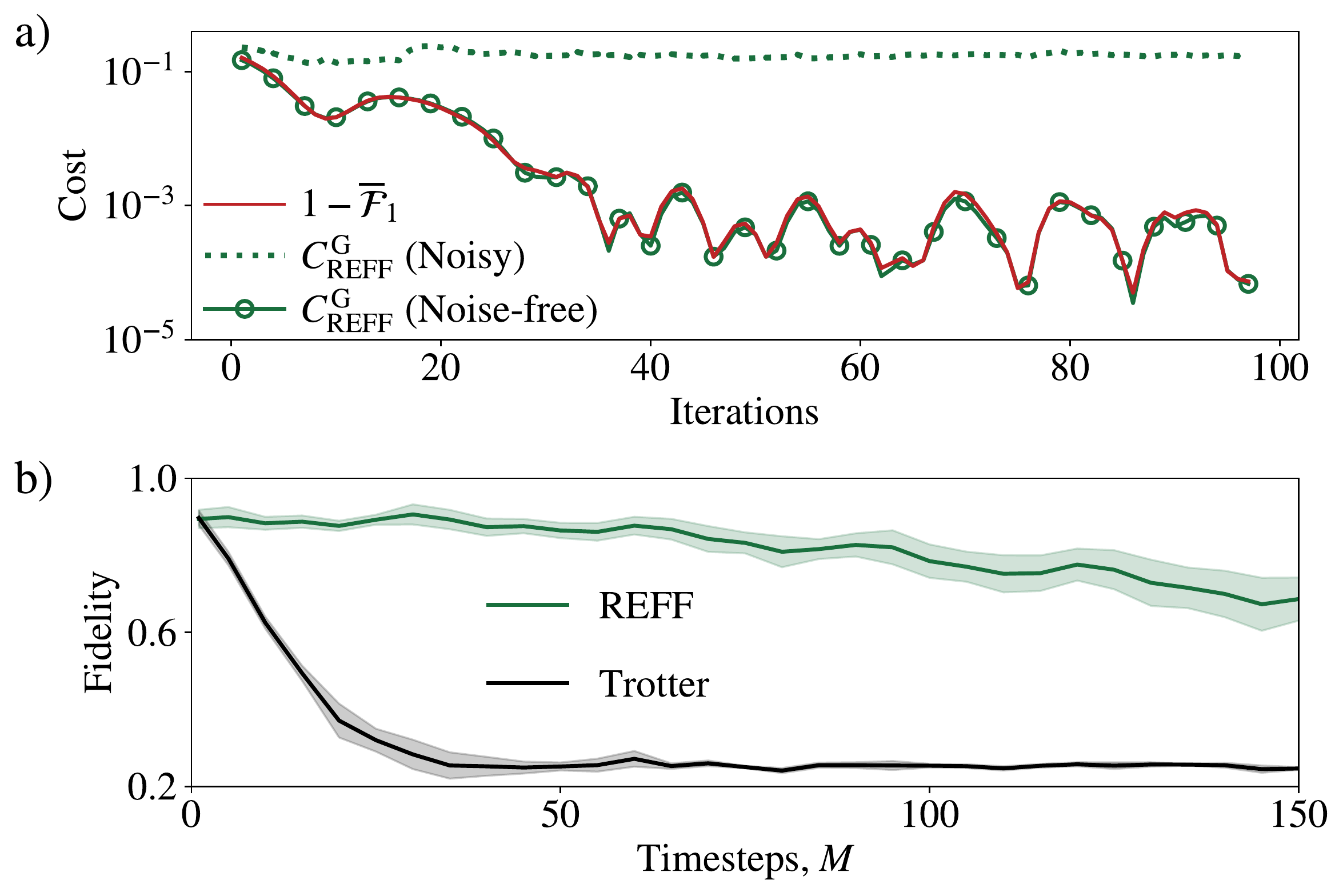}
\vspace{-6mm}
\caption{\small \textbf{Quantum hardware implementation.} a) The 2-qubit XY Hamiltonian is diagonalized via REFF on ibmq\_bogota using $2^{16}$ measurement shots per circuit. The noisy $C^{\rm G}_{\rm REFF}$ cost is measured on ibmq\_bogota, whereas the noise-free $C^{\rm G}_{\rm REFF}$ cost and $1-\overline{\mathcal{F}}_1$ are computed classically. b) After training, the fast-forwarded performance is compared to the iterated Trotter method for 5 Haar-random product states and we plot the mean and standard deviations.}
\label{fig:HardwareFastForwarding}
\end{figure}

To demonstrate the suitability of REFF for near-term hardware, we implemented REFF to diagonalize and fast-forward a 2-qubit spin chain, described by the XY Hamiltonian.
The clear alignment between $C^{\rm G}_{\rm REFF}$ and $1-\overline{\mathcal{F}}_1$ in Fig.~\ref{fig:HardwareFastForwarding}b) shows that, whilst only using a single unentangled training state, a full approximate diagonalization has been successfully learned. This then enables the high fidelity fast forwarding shown in Fig.~\ref{fig:HardwareFastForwarding}b). Namely, for random input states we achieved, on average, a fidelity of 0.8 for 94 time steps~\footnote{Quantum state tomography is used to reconstruct the output density matrix, $\rho$, of the noisy-fast-forwarding, which is used to compute the fast-forwarding fidelity, $\langle \psi | \rho | \psi \rangle$, where $|\psi\rangle$ is the target noise-free state.}. This is a factor of 23.8 improvement on the standard Trotter method which has a fidelity of less than 0.8 after 4 time steps. 

\medskip

\paragraph*{Discussion}

In this work, we introduced a framework for leveraging the power of QML for dynamical quantum simulations. The core idea is that quantum training data may be used to train a time-dependent QNN, which can then predict the evolution of the properties of the target system at long times using a short-depth circuit. By way of example, we introduced the REFF algorithm, which uses training data composed of product-state inputs and corresponding time-evolved outputs to learn an approximate diagonalization of the short-time evolution of the system. We showed that generalization bounds provide a tool to rigorously ground this QML-driven approach to quantum simulation. Specifically, for REFF we proved that a high fidelity simulation may be achieved with a number of training pairs $N$ that scales polynomially in the product of the number of fast forwarding steps, the number of parameters of the QNN, and the system size. 

While our error analysis and implementations focus on REFF, the framework is much more general. Important future steps include investigating alternative ans\"{a}tze for time-dependent QNNs and alternative forms of training data. The most appropriate choice of training data is dictated partially by the states and observables one wants to simulate and partially by what is available. For example, to simulate only within a particular subspace, one may use only training data from that subspace.
How to do so most efficiently remains an open question.
\medskip

\begin{acknowledgments}
We thank Sujay Kazi, Elliott Ball, and Robert M. Parrish for helpful conversations. ZH acknowledges support from the LANL Mark Kac Fellowship. MCC was supported by the TopMath Graduate Center of the TUM Graduate School at the Technical University of Munich, Germany, the TopMath Program at the Elite Network of Bavaria, by a doctoral scholarship of the German Academic Scholarship Foundation (Studienstiftung des deutschen Volkes), and by the BMWi (PlanQK).NE was supported by the U.S. DOE, Department of Energy Computational Science Graduate Fellowship under Award Number DE-SC0020347. HH is supported by a Google PhD Fellowship. PJC and ATS acknowledge initial support from the Los Alamos National Laboratory (LANL) ASC Beyond Moore's Law project. ATS was also supported by the Laboratory Directed Research and Development (LDRD) program of Los Alamos National Laboratory under project number 20210116DR. LC acknowledges support from LDRD program of LANL under project number 20200022DR. LC and PJC were also supported by the U.S. DOE, Office of Science, Office of Advanced Scientific Computing Research, under the Quantum Computing Application Teams (QCAT) program.
\end{acknowledgments}

\bibliography{quantum}

\clearpage 
\appendix

\setcounter{page}{1}
\renewcommand\thefigure{\thesection\arabic{figure}}
\setcounter{figure}{0} 

\onecolumngrid

\begin{center}
\large{ Supplementary Material for \\ ``Dynamical simulation via quantum machine learning with provable generalization''
}
\end{center}

\section{Generalization bounds for variational quantum machine learning and unitary compiling}\label{sct:generalization-bounds-prelims}

Here, we recall the theoretical results of Ref.~\cite{caro2021generalization} and discuss how they apply to our setting with quantum training data.
Mathematically, a \emph{quantum neural network (QNN)} (or \emph{variational quantum machine learning model}) is a parametrized completely positive and trace-preserving (CPTP) map $\mathcal{E}^{\mathrm{QNN}}_{\alv}(\cdot)$, where the parameter vector $\alv=(\vec{k}, \vec{\theta})$ consists of continuous parameters $\vec{\theta}$ and potentially also of discrete parameters $\vec{k}$. Importantly, the parametrization is determined by the structure of the QNN, e.g., by the layout of the underlying quantum circuit and the chosen parametrization of the gates therein.
The performance of such a QNN on a data point $(\ket{\Psi},\ket{\Phi})$ is evaluated by a cost/loss function
\begin{equation}
    \ell (\vec{\alpha};\ket{\Psi},\ket{\Phi})
    =\Tr\left[O^{\mathrm{loss}}_{\ket{\Psi},\ket{\Phi}}\mathcal{E}^{\mathrm{QNN}}_{\vec{\alpha}}(\ket{\Psi}\bra{\Psi})\right] \label{eq:loss-function} \, ,
\end{equation}
with a Hermitian loss observable $O^{\mathrm{loss}}_{\ket{\Psi},\ket{\Phi}}$, which may depend on the data point.

When using a QNN for machine learning purposes, given a data set $\mathcal{D}=\mathcal{D}(N) = \{(\ket{\Psi^{(j)}},\ket{\Phi^{(j)}}) \}_{j=1}^{N}$ consisting of $N$ input-output pairs, we train the parameters $\vec{\alpha}$ in such a way as to achieve small average cost on the training data. The latter is given as
\begin{equation}
    C_{\mathcal{D}(N)}(\vec{\alpha})
    = \frac{1}{N} \sum_{j=1}^N \ell (\vec{\alpha}; \ket{\Psi^{(j)}},\ket{\Phi^{(j)}})
\end{equation}
and is also often called the \emph{training error} of the QNN with parameter setting $\alv$.
To analyze the performance of the QNN beyond the training data, we consider the \emph{prediction error}, given by the expected cost
\begin{equation}
    C_P (\vec{\alpha})
    = \mathbb{E}_{(\ket{\Psi},\ket{\Phi})\sim P} \left[\ell (\vec{\alpha}; \ket{\Psi},\ket{\Phi})\right]\, ,
\end{equation}
where $P$ is the data-generating probability distribution.
In using the QNN, our goal is to identify a parameter setting that achieves small prediction error. However, as the data-generating distribution is typically unknown, we cannot directly evaluate this quantity. Instead, a common strategy in machine learning is to attempt to achieve small training error instead. Justifying this strategy requires bounding the \emph{generalization error} $\operatorname{gen}(\vec{\alpha})= C_P (\vec{\alpha}) - C_{\mathcal{D}(N)}(\vec{\alpha})$.

In this work, we make use of the following two generalization guarantees for QNNs. The first applies to QNNs that have only continuous parameters, but no discrete parameters allowing for training the structure of the QNN:

\begin{thm}[\cite{caro2021generalization}, Theorem $11$]\label{thm:generalization-bound-general}
    Let $\mathcal{E}^{\mathrm{QNN}}_{\vec{\theta}}(\cdot)$ be a QNN with a fixed architecture consisting of $K$ parameterized local CPTP maps and an arbitrary number of non-trainable, global CPTP maps. 
    Let $P$ be a probability distribution over input-output pairs. Suppose that, given training data $\mathcal{D}(N) = \{(\ket{\Psi^{(j)}},\ket{\Phi^{(j)}}) \}_{j=1}^{N}$ of size $N$, with the $(\ket{\Psi^{(j)}},\ket{\Phi^{(j)}})$ drawn i.i.d.~according to $P$, our optimization yields the parameter setting $\vec{\theta}^\ast = \vec{\theta}^\ast(\mathcal{D}(N))$.

    Then, with probability at least $1-\delta$ over the choice of i.i.d.~training data $\mathcal{D}(N)$ of size $N$ according to $P$, 
    \begin{equation}
        C_P (\vec{\theta}^\ast) - C_{\mathcal{D}(N)}(\vec{\theta}^\ast)
        \leq \mathcal{O}\left(C_\mathrm{loss} \left(\sqrt{\frac{K\log\left( K\right)}{N}} + \sqrt{\frac{\log(\nicefrac{1}{\delta})}{N}} \right)\right) \, ,\label{eq:gen-bound-fixed-architecture}
    \end{equation}
    with $C_\mathrm{loss} = \sup_{\ket{\Psi},\ket{\Phi}}\norm{O^{\mathrm{loss}}_{\ket{\Psi},\ket{\Phi}}}$.
\end{thm}

To also allow for optimizing over the structure of the quantum circuit used in the QNN, we use the following:

\begin{thm}[\cite{caro2021generalization}, Corollary $3$]\label{thm:generalization-bound-variable-architecture}
    Let $\mathcal{E}^{\mathrm{QNN}}_{\vec{\alpha}}(\cdot)$ be a QNN with a variable structure. Suppose that, for every $\kappa\in\mathbb{N}$, there are at most $G_\kappa\in\mathbb{N}$ allowed structures with exactly $\kappa$ parameterized local CPTP maps and an arbitrary number of non-trainable, global CPTP maps.
    Let $P$ be a probability distribution over input-output pairs. Suppose that, given training data $\mathcal{D}(N) = \{(\ket{\Psi^{(j)}},\ket{\Phi^{(j)}}) \}_{j=1}^{N}$ of size $N$, with the $(\ket{\Psi^{(j)}},\ket{\Phi^{(j)}})$ drawn i.i.d.~according to $P$, our optimization yields a (data-dependent) structure with $K=K(S)$ parameterized $2$-qubit CPTP maps and the parameter setting $\vec{\alpha}^\ast = \vec{\alpha}^\ast(\mathcal{D}(N))$.

    Then, with probability at least $1-\delta$ over the choice of i.i.d.~training data $\mathcal{D}(N)$ of size $N$ according to $P$,
    \begin{align}
        C_P (\vec{\alpha}^\ast) - C_{\mathcal{D}(N)}(\vec{\alpha}^\ast)
        \leq \mathcal{O}\left(C_\mathrm{loss} \left(\sqrt{\frac{K\log\left(K\right)}{N}} + \sqrt{\frac{\log G_K}{N}} + \sqrt{\frac{\log(\nicefrac{1}{\delta})}{N}} \right)\right)\, ,
    \end{align}
    with $C_\mathrm{loss} = \sup_{\ket{\Psi},\ket{\Phi}}\norm{O^{\mathrm{loss}}_{\ket{\Psi},\ket{\Phi}}}$.
\end{thm}

As observed in Ref.~\cite{caro2021generalization}, Theorems~\ref{thm:generalization-bound-general} and~\ref{thm:generalization-bound-variable-architecture} can be applied to obtain rigorous guarantees for variational unitary compilation. 
In this setting, training examples are of the form $(\ket{\Psi},\ket{\Phi})$, where $\ket{\Phi} = U\ket{\Psi}$ with $U$ the (unknown) target unitary to be learned.
The loss of a unitary QNN $\mathcal{V}^{\mathrm{QNN}}_{\vec{\alpha}}(\cdot) = V_{\vec{\alpha}}(\cdot)V_{\vec{\alpha}}^\dagger$ on the example $(\ket{\Psi},\ket{\Phi})$ is
\begin{align}
    \ell (\vec{\alpha};\ket{\Psi},\ket{\Phi})
    &= \frac{1}{4}\norm{\ket{\Phi}\bra{\Phi} - \mathcal{V}^\mathrm{QNN}_{\vec{\alpha}}(\ket{\Psi}\bra{\Psi})}_1^2\\
    &= 1 - \Tr\left[\ket{\Phi}\bra{\Phi}\cdot \mathcal{V}^\mathrm{QNN}_{\vec{\alpha}}(\ket{\Psi}\bra{\Psi})\right]\\
    &= 1 - \Tr\left[V_{\vec{\alpha}}^\dagger\ket{\Phi}\bra{\Phi}V_{\vec{\alpha}}\cdot \ket{\Psi}\bra{\Psi}\right]\, .
\end{align}
For convenience, we change the perspective to the Heisenberg picture. That is, given a training data point $(\ket{\Psi},\ket{\Phi})$, we think of $\ket{\Phi}$ as the input state to the unknown unitary $U^\dagger$ and of $\ket{\Psi}$ as the output state. Then, we see that the loss above is of the form of Eq.~\eqref{eq:loss-function}, with loss observable
\begin{align}
    O^{\mathrm{loss}}_{\ket{\Psi},\ket{\Phi}}
    &= \mathds{1} - \ket{\Psi}\bra{\Psi}\, ,
\end{align}
which gives
\begin{align}
    \ell (\vec{\alpha};\ket{\Psi},\ket{\Phi})
    &= \Tr\left[V_{\vec{\alpha}}^\dagger\ket{\Phi}\bra{\Phi}V_{\vec{\alpha}}\cdot O^{\mathrm{loss}}_{\ket{\Psi},\ket{\Phi}}\right]\, \label{eq:unitary-compiling-loss-function} .
\end{align}
Correspondingly, given a training data set $\mathcal{D}(N) = \{(\ket{\Psi^{(j)}},\ket{\Phi^{(j)}}) \}_{j=1}^{N}$ of size $N$, the training loss takes the form
\begin{align}
    C_{\mathcal{D}(N)}(\vec{\alpha})
    &= \frac{1}{N}\sum\limits_{j=1}^N \Tr\left[V_{\vec{\alpha}}^\dagger\ket{\Phi^{(j)}}\bra{\Phi^{(j)}}V_{\vec{\alpha}}\cdot O^{\mathrm{loss}}_{\ket{\Psi^{(j)}},\ket{\Phi^{(j)}}}\right]\label{eq:unitary-compiling-cost-general}\\
    &= \frac{1}{N}\sum\limits_{j=1}^N \left( 1 - \left\lvert\bra{\Phi^{(j)}} V_{\vec{\alpha}}\ket{\Psi^{(j)}}\right\rvert^2\right)\, .
\end{align}
The expected cost is given by
\begin{align}
    C_P (\vec{\alpha})
    &= \mathbb{E}_{\ket{\Psi}\sim P}\left[\Tr\left[V_{\vec{\alpha}}^\dagger\ket{\Phi}\bra{\Phi}V_{\vec{\alpha}}\cdot O^{\mathrm{loss}}_{\ket{\Psi},\ket{\Phi}}\right]\right]\label{eq:unitary-compiling-expected-cost-general}\\
    &= 1 - \mathbb{E}_{\ket{\Psi}\sim P}\left[\left\lvert\bra{\Phi}V_{\vec{\alpha}}\ket{\Psi}\right\rvert^2\right]\, ,
\end{align}
where $\ket{\Phi}=U\ket{\Psi}$ and $P$ is the data-generating measure. In this setting, the guarantees of Theorems~\ref{thm:generalization-bound-general} and~\ref{thm:generalization-bound-variable-architecture} become: 

\begin{corol}\label{corol:generalization-bound-unitary-compilation-general}
    Let $\mathcal{V}^{\mathrm{QNN}}_{\vec{\theta}}(\cdot)$ be a unitary QNN with a fixed architecture consisting of $K$ parameterized local unitaries and an arbitrary number of non-trainable, global unitaries. 
    Let $P$ be a probability distribution over input-output pairs. Suppose that, given training data $\mathcal{D}(N) = \{(\ket{\Psi^{(j)}},\ket{\Phi^{(j)}}) \}_{j=1}^{N}$ of size $N$, with the $\ket{\Psi^{(j)}}$ drawn i.i.d.~according to $P$ and $\ket{\Phi^{(j)}} = U\ket{\Psi^{(j)}}$, our optimization yields the parameter setting $\vec{\theta}^\ast = \vec{\theta}^\ast(\mathcal{D}(N))$.

    Then, with probability at least $1-\delta$ over the choice of i.i.d.~training data $\mathcal{D}(N)$ of size $N$ according to $P$ and $U$, 
    \begin{equation}
        C_P (\vec{\theta}^\ast) - C_{\mathcal{D}(N)}(\vec{\theta}^\ast)
        \leq \mathcal{O}\left(\sqrt{\frac{K\log\left(K\right)}{N}} + \sqrt{\frac{\log(\nicefrac{1}{\delta})}{N}} \right) \, .
    \end{equation}
\end{corol}

Again, this result has a variant for variable structure QNNs:

\begin{corol}\label{corol:generalization-bound-unitary-compilation-variable-structure}
    Let $\mathcal{V}^{\mathrm{QNN}}_{\vec{\alpha}}(\cdot)$ be a unitary QNN with a variable structure. Suppose that, for every $\kappa\in\mathbb{N}$, there are at most $G_\kappa\in\mathbb{N}$ allowed structures with exactly $\kappa$ parameterized local unitaries and an arbitrary number of non-trainable, global unitaries.
    Let $P$ be a probability distribution over input-output pairs. Suppose that, given training data $\mathcal{D}(N) = \{(\ket{\Psi^{(j)}},\ket{\Phi^{(j)}}) \}_{j=1}^{N}$ of size $N$, with the $\ket{\Psi^{(j)}}$ drawn i.i.d.~according to $P$ and $\ket{\Phi^{(j)}} = U\ket{\Psi^{(j)}}$, our optimization yields a (data-dependent) structure with $K=K(S)$ parameterized local unitaries maps and the parameter setting $\vec{\alpha}^\ast = \vec{\alpha}^\ast(\mathcal{D}(N))$.

    Then, with probability at least $1-\delta$ over the choice of i.i.d.~training data $\mathcal{D}(N)$ of size $N$ according to $P$ and $U$,
    \begin{align}
        C_P (\vec{\alpha}^\ast) - C_{\mathcal{D}(N)}(\vec{\alpha}^\ast)
        \leq \mathcal{O}\left(\sqrt{\frac{K\log\left(K\right)}{N}} + \sqrt{\frac{\log G_K}{N}} + \sqrt{\frac{\log(\nicefrac{1}{\delta})}{N}} \right)\, .
    \end{align}
\end{corol}

Note that we can define the loss function, and thus the training and expected cost, also using different loss observables $O^{\mathrm{loss}}_{\ket{\Psi^{(j)}},\ket{\Phi^{(j)}}}$ in Eqs.~\eqref{eq:unitary-compiling-loss-function},~\eqref{eq:unitary-compiling-cost-general}, and~\eqref{eq:unitary-compiling-expected-cost-general}. 
As Theorems~\ref{thm:generalization-bound-general} and~\ref{thm:generalization-bound-variable-architecture} allow for such general loss observables, a change of this kind does not alter the statements of Corollaries~\ref{corol:generalization-bound-unitary-compilation-general} and~\ref{corol:generalization-bound-unitary-compilation-variable-structure}.
We will make use of this observation in Section~\ref{sbsct:product-costs} to define a suitable local cost.

\section{Proofs of results presented in the main text}\label{Ap:Proofs}

In this appendix, we give complete proofs for all analytical results presented in the main text. We start in Section~\ref{sbsct:entangled-costs} by providing the error analysis for training using fully Haar random $n$-qubit training states. These training states will typically be highly entangled and so hard to prepare on near term hardware. 
While we do not advocate this training strategy in practice, the error analysis for this case is simpler and thus acts as preparation for our subsequent analysis. 
In Section~\ref{sbsct:product-costs}, we provide the error analysis for the approach proposed in the main text whereby the training data is composed of tensor products of locally random states.

\medskip 

\subsection{Training with Haar-random $n$-qubit input states}\label{sbsct:entangled-costs}

Haar-random $n$-qubit states provide a natural ensemble for generating data on which to train or test our ansatz $V_{\Delta t} ( \alv )$ for the short-time evolution.
When working in the setting of unitary compilation described in Appendix~\ref{sct:generalization-bounds-prelims}, choosing the $n$-qubit Haar measure as data-generating distribution in Eq.~\eqref{eq:unitary-compiling-cost-general} leads to training data of the form 
\begin{align}
    \mathcal{D}_{\rm E}(N)
    = \{(\ket{\Psi^{(j)}},\ket{\Phi^{(j)}}) \}_{j=1}^{N}
    = \left\{\left(\ket{\Psi^{(j)}}, U \ket{\Psi^{(j)}}\right)\right\}_{j=1}^N \, ,
\end{align}
where the $\ket{\Psi^{(j)}}$ are i.i.d.~Haar-random $n$-qubit states, typically highly entangled, and the $\ket{\Phi^{(j)}} = U \ket{\Psi^{(j)}}$ are the corresponding output states under the (approximate short-time) evolution $U=U_{\Delta t}$ that is to be learned. 
Plugging such a data set $\mathcal{D}_{\rm E}(N)$ into Eq.~\eqref{eq:unitary-compiling-cost-general}, we obtain the corresponding training cost
\begin{align}
    C_{\mathcal{D}_{\rm E}(N)}^{G}(\alv)
    &= \frac{1}{N}\sum\limits_{j=1}^N \Tr\left[V_{\vec{\alpha}}^\dagger\ket{\Phi^{(j)}}\bra{\Phi^{(j)}}V_{\vec{\alpha}}\cdot O^{\mathrm{loss}}_{\ket{\Psi^{(j)}},\ket{\Phi^{(j)}}}\right]\\
    &= 1 - \frac{1}{N}\sum\limits_{j=1}^N \lvert\bra{\Psi^{(j)}}U^\dagger V_{\Delta t} ( \alv )\ket{\Psi^{(j)}}\rvert^2 \, .
\end{align}
According to Eq.~\eqref{eq:unitary-compiling-expected-cost-general}, the corresponding expected cost is
\begin{align}
    C_{{\rm Haar}_n}^{G}(\alv) 
    &= \mathbb{E}_{\ket{\Psi} \sim {\rm Haar}_n} \left[\Tr\left[ V_{\Delta t} ( \alv )^\dagger\ket{\Phi}\bra{\Phi}V_{\Delta t} ( \alv )\cdot O^{G}_{\ket{\Psi},\ket{\Phi}}\right] \right] \label{eq:entangled-global-cost}\\
    &= 1 - \mathbb{E}_{\ket{\Psi} \sim {\rm Haar}_n} \left[ \Tr\left[ U \ket{\Psi}\bra{\Psi} U^\dagger V_{\Delta t} ( \alv ) \ket{\Psi}\bra{\Psi} V_{\Delta t} ( \alv )^\dagger \right] \right] \, ,
\end{align}
where the expectation is w.r.t.~states drawn from the $n$-qubit Haar measure, and $\ket{\Phi}=U\ket{\Psi}$, for $U$ the approximate short-time evolution unitary to be learned .

There is another natural candidate for measuring the distance between two unitaries, and thus for evaluating the cost in our compilation task. Namely, we could also consider the \emph{Hilbert-Schmidt test (HST) cost}, arising from the Hilbert-Schmidt inner product, given by
\begin{align}
    C_{\mbox {\tiny HST}}(U, V_{\Delta t} ( \alv )) = 1-\frac{1}{d^2}|\Tr[ U^\dagger V_{\Delta t} ( \alv )]|^2 \, , \label{eq:hst-cost}
\end{align}
where we write $d=2^n$. This cost was proposed for the variational compilation of unitaries in Ref.~\cite{khatri2019quantum}. In fact, the HST cost and the cost based on an expected fidelity over the $n$-qubit Haar measure are closely related:

\begin{lem}\label{lem:entangled-global-versus-average-fidelity-versus-hst}
    When taking a Haar-random $n$-qubit state as test state, the true cost for the time step $\Delta t$ takes the form
    \begin{equation}
        C_{{\rm Haar}_n}^{G}(\alv)
        = 1- \overline{F}(U, V_{\Delta t} ( \alv ))
        = \frac{d}{d+1} C_{\mbox {\tiny \rm HST}}(U, V_{\Delta t} ( \alv )) \, .
    \end{equation}
    Here, we denote by 
    \begin{equation}
        \overline{F}(U, V_{\Delta t} ( \alv ))
        \coloneqq \mathbb{E}_{\ket{\Psi} \sim {\rm Haar}_n} \left[\lvert\bra{\Psi}U^\dagger V_{\Delta t} ( \alv )\ket{\Psi}\rvert^2\right]
    \end{equation}
    the average fidelity over Haar-random inputs.
\end{lem}

\begin{proof}
    By definition of the cost function $C_{{\rm Haar}_n}^{G}(\alv)$ and the average fidelity $\overline{F}(U, V_{\Delta t} ( \alv ))$, we directly have $C_{{\rm Haar}_n}^{G}(\alv) = 1-\overline{F}(U, V_{\Delta t} ( \alv ))$. The relation
    \begin{align}
        C_{\mbox {\tiny \rm HST}}(U, V)
        &= 1-\frac{1}{d^2}|\Tr[U^\dagger V]|^2
        = \frac{d+1}{d}\left(1-\overline{F}(U, V)\right)
    \end{align}
    can, e.g., be found in Refs.~\cite{horodecki1999general, nielsen2002simple, khatri2019quantum}.
\end{proof}

Lemma~\ref{lem:entangled-global-versus-average-fidelity-versus-hst} effectively allows us to freely choose in our analysis whether we work with $C_{{\rm Haar}_n}^{G}(\alv)$, $\overline{F}(U, V_{\Delta t} ( \alv ))$, or $C_{\mbox {\tiny \rm HST}}(U, V_{\Delta t} ( \alv ))$. With this freedom, we can now establish a guarantee on the overall simulation fidelity for an evolution along a Hamiltonian $H$ for time $T$, given by
\begin{align}
    \overline{F}(\alv_{\rm opt}, T) 
    &=  \overline{F} (e^{-iTH}, V_T (\alv_{\rm opt})) \, , 
\end{align}
when using a QNN with $K$ trainable local gates to learn the approximate short-time evolution.
We denote by $M$ the number of time steps of size $\Delta t$ needed to achieve the overall time $T$, i.e., $T = M\cdot\Delta t$.
Moreover, as in the main text, we denote by 
\begin{equation}
    \epsilon = \lVert U_{\Delta t} - e^{- i H \Delta t}\rVert_2  \, ,
\end{equation}
the short-time approximation error (measured according to the Schatten $2$-norm) incurred by approximating $e^{- i H \Delta t}$ with $U_{\Delta t}$.
With this notation in place, we state and prove our first guarantee:

\begin{thm}[Simulation error (entangled training, global cost)]\label{thm:entangled-training-global-cost}
    For a QNN $V_{\Delta t} ( \alv )$ with $K$ parameterized local gates, with probability $\geq 1-\delta$ over the choice of training data $\mathcal{D}_{\rm E}(N)=\left\{\left(\ket{\Psi_i}, U \ket{\Psi_i}\right)\right\}_{i=1}^N$, with the $\ket{\Psi_i}$ independent Haar-random $n$-qubit states and $U=U_{\Delta t}$ the unknown approximate short-time evolution unitary, the total simulation fidelity satisfies
    \begin{align}
        \overline{F}(\alv_{\rm opt}, M\cdot\Delta t)
        &\geq 1 - 2M^2 \left(\frac{\epsilon^2}{d+1} + 2C_{\mathcal{D}_{\rm E}(N)}^{G}(\alv_{\rm opt})\right) - \mathcal{O}\left(M^2 \left(\sqrt{\frac{K\log\left(K\right)}{N}} + \sqrt{\frac{\log(\nicefrac{1}{\delta})}{N}} \right)\right) \, ,
    \end{align}
    where $\alv_{\rm opt}$ denotes the final parameter setting after training.
\end{thm}

\begin{proof}
    By Theorem~\ref{thm:generalization-bound-general}, with probability $\geq 1-\delta$ over the choice of training data $\mathcal{D}_{\rm E}(N)=\left\{\left(\ket{\Psi_i}, U \ket{\Psi_i}\right)\right\}_{i=1}^N$, with the $\ket{\Psi_i}$ independent Haar-random $n$-qubit states, the final parameter setting $\alv_{\rm opt}$ satisfies
    \begin{equation}
        C_{{\rm Haar}_n}^{G}(\alv_{\rm opt})
        \leq C_{\mathcal{D}_{\rm E}(N)}^{G}(\alv_{\rm opt}) + \mathcal{O} \left(\sqrt{\frac{K\log\left(K\right)}{N}} + \sqrt{\frac{\log(\nicefrac{1}{\delta})}{N}} \right)\, .
    \end{equation}
    Once we recall from Lemma~\ref{lem:entangled-global-versus-average-fidelity-versus-hst} that $C_{{\rm Haar}_n}^{G}(\alv_{\rm opt}) = \frac{d}{d+1} C_{\mbox {\tiny \rm HST}}(U, V_{\Delta t} (\alv_{\rm opt}))$, we see that the above gives a bound on the HST cost of the learned short-time evolution in terms of the training cost on $N$ Haar-random $n$-qubit states, which holds with high probability. Namely, with probability $\geq 1-\delta$, 
    \begin{equation}
        C_{\mbox {\tiny \rm HST}}(U, V_{\Delta t} (\alv_{\rm opt}))
        \leq \frac{d+1}{d}C_{\mathcal{D}_{\rm E}(N)}^{G}(\alv_{\rm opt}) + \mathcal{O} \left(\frac{d+1}{d}\left(\sqrt{\frac{K\log\left(K\right)}{N}} + \sqrt{\frac{\log(\nicefrac{1}{\delta})}{N}} \right)\right)\, . \label{eq:generalization-bound-hst-cost-short-time}
    \end{equation}
    
    Next, we make use of the behavior of the HST cost under iterated applications of unitaries, which it inherits from the Schatten $2$-norm. 
    Namely, according to Eq.~(S$20$) of Ref.~\cite{cirstoiu2020variational}, we have
    \begin{equation}
        \sqrt{1-\sqrt{1-\frac{d+1}{d}(1-\overline{F}(\alv_{\rm opt}, M\cdot\Delta t))}} 
        \leq  \frac{M}{\sqrt{2d}} \epsilon + M \sqrt{1-\sqrt{1-C_{\mbox {\tiny \rm HST}}(U, V_{\Delta t} (\alv_{\rm opt}) }} \, . 
    \end{equation}
    Rearranging this inequality, and applying both $\sqrt{1-x}\geq 1-x$ for $x\in [0,1]$ and $(a+b)^2\leq 2(a^2+b^2)$ for $a,b\in\mathbb{R}$, we can conclude
    \begin{equation}
        \overline{F}(\alv_{\rm opt}, M\cdot\Delta t)
        \geq  1 - \frac{2M^2}{d+1}\left[\epsilon^2 + 2d\cdot C_{\mbox {\tiny \rm HST}}(U, V_{\Delta t} (\alv_{\rm opt}))\right] \, , \label{eq:overall-simulation-fidelity-bound-via-hst-cost}
    \end{equation}
    where $\epsilon$ denotes the short-time approximation error as defined above.
    Now, we can plug in the bound from Eq.~\eqref{eq:generalization-bound-hst-cost-short-time} and obtain
    \begin{align}
        \overline{F}(\alv_{\rm opt}, M\cdot\Delta t)
        &\geq  1 - \frac{2M^2}{d+1}\left[\epsilon^2 + 2d\cdot \left(\frac{d+1}{d}C_{\mathcal{D}_{\rm E}(N)}^{G}(\alv_{\rm opt}) + \mathcal{O} \left(\frac{d+1}{d}\left(\sqrt{\frac{K\log\left(K\right)}{N}} + \sqrt{\frac{\log(\nicefrac{1}{\delta})}{N}} \right)\right)\right)\right]\\
        &= 1 - 2M^2 \left(\frac{\epsilon^2}{d+1} + 2C_{\mathcal{D}_{\rm E}(N)}^{G}(\alv_{\rm opt})\right) - \mathcal{O}\left(M^2 \left(\sqrt{\frac{K\log\left(K\right)}{N}} + \sqrt{\frac{\log(\nicefrac{1}{\delta})}{N}} \right)\right) \, ,
    \end{align}
    which is the claimed bound.
\end{proof}

Theorem~\ref{thm:entangled-training-global-cost} tells us: When using a QNN to learn the approximate short-time evolution by training on a data set with Haar random input states, a training data size effectively scaling as $N\sim M^4 \cdot K\log(K)$ will, with high probability, lead to good generalization also on the level of the long-time evolution. 

\begin{rmk}\label{rmk:termination-condition-theory-global}
    From a more practical perspective, we can interpret Theorem~\ref{thm:entangled-training-global-cost} as justifying a purely training-cost-based termination condition for training if a large enough training data set is used.
    We can see this as follows:
    Suppose we want to achieve a simulation fidelity $\geq 1-\varepsilon$ for up to $M_0$ forwarding steps, with some fixed success probability, say $\geq 0.95$.
    Then, it suffices to choose training data of size $N\sim M_0^4\cdot \tfrac{K\log (K)}{(\varepsilon/2)^2}$ and to ensure that the training cost and the short-time approximation error satisfy $\tfrac{\epsilon^2}{d+1} + 2C_{\mathcal{D}_{\rm E}(N)}^{G}(\alv_{\rm opt})\lesssim \tfrac{\varepsilon}{4M_0^2}$.
    In other words, assuming a suitable training data size is chosen, we can terminate training as soon as the training cost satisfies $C_{\mathcal{D}_{\rm E}(N)}^{G}(\alv_{\rm opt})\lesssim \tfrac{\varepsilon}{8M_0^2} - \tfrac{\epsilon^2}{2(d+1)}$.
    (Naturally, this presupposes a small enough short-time approximation error, namely $\tfrac{\epsilon^2}{d+1}< \tfrac{\varepsilon}{4M_0^2}$.)
\end{rmk}

\begin{rmk}
    Throughout our proofs in this subsection, the only property of the $n$-qubit Haar measure ${\rm Haar}_n$ that entered our reasoning was the connection of the corresponding expected cost to the average fidelity and the HST cost, see Lemma~\ref{lem:entangled-global-versus-average-fidelity-versus-hst}.
    We can replace the data-generating measure ${\rm Haar}_n$ by an $n$-qubit $2$-design. This will lead to the same expected cost, still satisfying Lemma~\ref{lem:entangled-global-versus-average-fidelity-versus-hst}, and thus to the same simulation error guarantee.
\end{rmk}

\subsection{Training with product of Haar-random single-qubit input states}\label{sbsct:product-costs}

The costs presented in Section~\ref{sbsct:entangled-costs} are based on Haar-random $n$-qubit states. 
However, these are costly to prepare in practice as they require deep quantum circuits.
In particular, assuming access to a training data set consisting of multiple Haar-random $n$-qubit states and their output states under (approximate) short-time evolution might be too optimistic for many practical applications as $n$ grows. Therefore, we instead use a different notion of cost, based on tensor products of easy-to-prepare Haar-random single-qubit states. 
That is, we work with training data of the form 
\begin{align}
    \mathcal{D}_{\rm P}(N)
    = \{(\ket{\Psi_{\rm P}^{(j)}},\ket{\Phi^{(j)}_{\rm P}}) \}_{j=1}^{N}
    = \left\{\left(\bigotimes_{i=1}^n \ket{\psi^{(j)}_i}, U\left(\bigotimes_{i=1}^n \ket{\psi^{(j)}_i}\right)\right)\right\}_{j=1}^N \, ,
\end{align}
where the $\ket{\psi^{(j)}_i}$ are drawn i.i.d.~according to the single qubit Haar measure ${\rm Haar}_1$ and $U = U_{\Delta t}$ is the unknown short time evolution that we are trying to learn. 
Note that, if we have access to a black box implementing the evolution according to $U$, we can create such a training data set efficiently because only local Haar random states are required as inputs.

\subsubsection{Global cost function}

If we plug this training data for product state inputs into Eq.~\eqref{eq:unitary-compiling-cost-general}, we see that the training cost is given by
\begin{align}\label{eq:globalprod}
    C_{ \mathcal{D}_{\rm P}(N)}^{G}(\alv)
    &= \frac{1}{N}\sum\limits_{j=1}^N \Tr\left[V_{\vec{\alpha}}^\dagger\ket{\Phi^{(j)}_{\rm P}}\bra{\Phi^{(j)}_{\rm P}}V_{\vec{\alpha}}\cdot O^{G}_{\ket{\Psi_{\rm P}}, \ket{\Phi_{\rm P}}}\right]\\
    &= 1 -\frac{1}{N}\sum\limits_{j=1}^N \Tr \left[ U \ket{\Psi_{\rm P}^{(j)}}\bra{\Psi_{\rm P}^{(j)}}  U^\dagger V_{\Delta t} ( \alv )\ket{\Psi_{\rm P}^{(j)}}\bra{\Psi_{\rm P}^{(j)}} V_{\Delta t} ( \alv )^\dagger\right] \, .
\end{align}
This cost is a global cost because the loss observables $O^{\mathrm{loss}}_{\ket{\Psi_{\rm P}^{(j)}}, \ket{\Phi_{\rm P}^{(j)}}} = O^{G}_{\ket{\Psi_{\rm P}^{(j)}}, \ket{\Phi_{\rm P}^{(j)}}} = \mathds{1} - \ket{\Psi_{\rm P}^{(j)}}\bra{\Psi_{\rm P}^{(j)}}$ are global measurements. That is, they require performing measurements on all $n$ qubits.
We will begin by providing rigorous results for the global cost.
Later, we will discuss generalization to the local version that enjoys a better trainability.

Following Eq.~\eqref{eq:unitary-compiling-expected-cost-general}, for the global cost using product training states, Eq.~\eqref{eq:globalprod}, the expected cost is
\begin{align}
    C_{{\rm Haar}_1^{\otimes n}}^{G}(\alv) 
    &= \mathbb{E}_{\ket{\Psi_{\rm P}} \sim {\rm Haar}_1^{\otimes n}} \left[\Tr\left[ V_{\Delta t} ( \alv )^\dagger\ket{\Phi_{\rm P}}\bra{\Phi_{\rm P}}V_{\Delta t} ( \alv )\cdot O^{G}_{\ket{\Psi_{\rm P}},\ket{\Phi_{\rm P}}}\right] \right] \label{eq:product-global-cost}\\
    &= 1 - \mathbb{E}_{\ket{\Psi_{\rm P}} \sim {\rm Haar}_1^{\otimes n}}  \left[ \Tr \left[ U \ket{\Psi_{\rm P}}\bra{\Psi_{\rm P}}  U^\dagger V_{\Delta t} ( \alv )\ket{\Psi_{\rm P}}\bra{\Psi_{\rm P}} V_{\Delta t} ( \alv )^\dagger\right] \right] \, .
\end{align}
Here, we write $\Psi_{\rm P} \coloneqq \bigotimes_{i=1}^n \ket{\psi_i} \sim {\rm Haar}_1^{\otimes n}$ to express that the expectation is over tensor products of states, each drawn from ${\rm Haar}_1$, and $\ket{\Phi_{\rm P}} = U \ket{\Psi_{\rm P}}$.

As a special case of~\cite[Lemma $1$]{caro2022outofdistribution}, we obtain that the product cost $C_{{\rm Haar}_1^{\otimes n}}^{G}(\alv)$ is closely related to the entangled cost $C_{{\rm Haar}}^{G}(\alv)$:

\begin{lem}[Product versus entangled cost (global)]\label{lem:product-versus-entangled-cost}
        For any parameter setting $\alv$, we have the following relation between the global versions of the product and the entangled cost on $n$ qubits:
    \begin{align}
        C_{{\rm Haar}_1^{\otimes n}}^{G}(\alv)
        &\leq  \frac{d+1}{d} C_{{\rm Haar}_n}^{G}(\alv)
        \leq 2 C_{{\rm Haar}_1^{\otimes n}}^{G}(\alv) \, .
    \end{align}
\end{lem}

With this connection between $C_{{\rm Haar}_n}^{G}(\alv)$ and $C_{{\rm Haar}_1^{\otimes n}}^{G}(\alv)$ in place, we can now prove the first part of Theorem~\ref{thm:product-training-global-cost}, which we state formally here:

\begin{thm}[Simulation error (product training, global cost) -- First half of Theorem~\ref{thm:product-training-global-cost}]\label{thm:product-training-global-cost-appendix}
    For a QNN $V_{\Delta t} ( \alv )$ with $K$ parameterized local gates, with probability $\geq 1-\delta$ over the choice of training data $\mathcal{D}_{\rm P}(N)=\left\{ \bigotimes_{j=1}^n \ket{\psi_i^{(j)}}, U \bigotimes_{j=1}^n \ket{\psi_i^{(j)}} \right\}_{i=1}^N$, with the $\ket{\psi_i^{(j)}}$ independent Haar-random single-qubit states, the total simulation fidelity satisfies
    \begin{align}
        \overline{F}(\alv_{\rm opt}, M\cdot\Delta t)
        &\geq 1 - 2M^2\left[\frac{\epsilon^2}{d+1} + 4 C_{\mathcal{D}_{\rm P}(N)}^{G}(\alv_{\rm opt})\right]
        - \mathcal{O} \left(M^2\left(\sqrt{\frac{K\log\left(K\right)}{N}}+ \sqrt{\frac{\log(\nicefrac{1}{\delta})}{N}} \right)\right) \, ,
    \end{align}
    where $\alv_{\rm opt} = \alv_{\rm opt} (\mathcal{D}_{\rm P}(N))$ denotes the final parameter setting after training.
\end{thm}

\begin{proof}
    By Theorem~\ref{thm:generalization-bound-general} and Corollary~\ref{corol:generalization-bound-unitary-compilation-general}, we know: With probability $\geq 1-\delta$ over the choice of training data $\mathcal{D}_{\rm P}(N)=\left\{ \bigotimes_{j=1}^n \ket{\psi_i^{(j)}}, U \bigotimes_{j=1}^n \ket{\psi_i^{(j)}} \right\}_{i=1}^N$, with the $\ket{\psi_i^{(j)}}$ independent Haar-random single-qubit states, the final parameter setting $\alv_{\rm opt}=\alv_{\rm opt}(\mathcal{D}_{\rm P}(N))$ satisfies
    \begin{align}
        C_{{\rm Haar}_1^{\otimes n}}^{G}(\alv) 
        &\leq C_{\mathcal{D}_{\rm P}(N)}^{G}(\alv_{\rm opt}) + \mathcal{O} \left(\sqrt{\frac{K\log\left(K\right)}{N}} + \sqrt{\frac{\log(\nicefrac{1}{\delta})}{N}} \right)\, .
    \end{align}
    By Lemmas~\ref{lem:entangled-global-versus-average-fidelity-versus-hst} and Lemma~\ref{lem:product-versus-entangled-cost}, this implies that, with probability $\geq 1-\delta$, 
    \begin{equation}
        C_{\mbox {\tiny \rm HST}}(U, V_{\Delta t} (\alv_{\rm opt}))
        =\frac{d+1}{d}C_{{\rm Haar}_n}^{G}(\alv)
        \leq 2 C_{{\rm Haar}_1^{\otimes n}}^{G}(\alv)
        \leq 2 C_{\mathcal{D}_{\rm P}(N)}^{G}(\alv_{\rm opt}) +\mathcal{O} \left(\sqrt{\frac{K\log\left(K\right)}{N}}+ \sqrt{\frac{\log(\nicefrac{1}{\delta})}{N}} \right) \, .
    \end{equation}
    We now recall from Eq.~\eqref{eq:overall-simulation-fidelity-bound-via-hst-cost} that
    \begin{equation}
        \overline{F}(\alv_{\rm opt}, M\cdot\Delta t)
        \geq  1 - \frac{2M^2}{d+1}\left[\epsilon^2 + 2d\cdot C_{\mbox {\tiny \rm HST}}(U, V_{\Delta t} (\alv_{\rm opt}))\right] \, ,
    \end{equation}
    with $\epsilon$ the short-time approximation error 
    \begin{equation}
        \epsilon = \lVert U_{\Delta t} - e^{- i H \Delta t}\rVert_2  \, .
    \end{equation}
    If we plug our above upper bound on the HST cost $C_{\mbox {\tiny \rm HST}}(U, V_{\Delta t} (\alv_{\rm opt}))$ into this lower bound for the overall simulation fidelity $\overline{F}(\alv_{\rm opt}, M\cdot\Delta t)$, we obtain: With probability $\geq 1-\delta$,
    \begin{align}
        \overline{F}(\alv_{\rm opt}, M\cdot\Delta t)
        &\geq  1 - \frac{2M^2}{d+1}\left[\epsilon^2 + 2d\cdot \left( 2C_{\mathcal{D}_{\rm P}(N)}^{G}(\alv_{\rm opt}) + \mathcal{O} \left(\sqrt{\frac{K\log\left(K\right)}{N}}+ \sqrt{\frac{\log(\nicefrac{1}{\delta})}{N}} \right)\right)\right]\\
        &= 1 - 2M^2\left[\frac{\epsilon^2}{d+1} + \frac{4d}{d+1}\cdot C_{\mathcal{D}_{\rm P}(N)}^{G}(\alv_{\rm opt}) \right]
        - \mathcal{O} \left(M^2\cdot\frac{d}{d+1}\cdot\left(\sqrt{\frac{K\log\left(K\right)}{N}}+ \sqrt{\frac{\log(\nicefrac{1}{\delta})}{N}} \right)\right)\\
        &\geq 1 - 2M^2\left[\frac{\epsilon^2}{d+1} + 4 C_{\mathcal{D}_{\rm P}(N)}^{G}(\alv_{\rm opt})\right]
        - \mathcal{O} \left(M^2\left(\sqrt{\frac{K\log\left(K\right)}{N}}+ \sqrt{\frac{\log(\nicefrac{1}{\delta})}{N}} \right)\right)\, ,
    \end{align}
    as claimed.
\end{proof}

In the case of a variable structure QNN for learning the approximate short-time evolution, we have the following guarantee:

\begin{corol}\label{corol:product-training-global-cost-variable-structure}
    For a QNN $V_{\Delta t} ( \alv )$ with a variable structure such that, for every $\kappa\in\mathbb{N}$, there are at most $G_\kappa\in\mathbb{N}$ allowed structures with exactly $\kappa$ parameterized local gates, with probability $\geq 1-\delta$ over the choice of training data $\mathcal{D}_{\rm P}(N)=\left\{ \bigotimes_{j=1}^n \ket{\psi_i^{(j)}}, U \bigotimes_{j=1}^n \ket{\psi_i^{(j)}} \right\}_{i=1}^N$, with the $\ket{\psi_i^{(j)}}$ independent Haar-random single-qubit states, the total simulation fidelity satisfies
    \begin{align}
        \overline{F}(\alv_{\rm opt}, M\cdot\Delta t)
        &\geq 1 - 2M^2\left[\frac{\epsilon^2}{d+1} + 4 C_{\mathcal{D}_{\rm P}(N)}^{G}(\alv_{\rm opt}) \right]
        - \mathcal{O} \left(M^2\left(\sqrt{\frac{K\log\left(K\right)}{N}}+ \sqrt{\frac{\log G_K}{N}} + \sqrt{\frac{\log(\nicefrac{1}{\delta})}{N}} \right)\right) \, ,
    \end{align}
    where $K=K(\mathcal{D}_{\rm P}(N))$ denotes the number of parameterized local gates used in the final structure after training and $\alv_{\rm opt} = \alv_{\rm opt} (\mathcal{D}_{\rm P}(N))$ denotes the final parameter setting after training.
\end{corol}

\begin{proof}
    The proof of this Corollary is analogous to the proof of Theorem~\ref{thm:product-training-global-cost-appendix}, we only have to use the generalization guarantees of Theorem~\ref{thm:generalization-bound-variable-architecture} and Corollary~\ref{corol:generalization-bound-unitary-compilation-variable-structure} instead of Theorem~\ref{thm:generalization-bound-general} and Corollary~\ref{corol:generalization-bound-unitary-compilation-general}.
\end{proof}

Analogously to Remark~\ref{rmk:termination-condition-theory-global}, one can use Theorem~\ref{thm:product-training-global-cost-appendix} and Corollary~\ref{corol:product-training-global-cost-variable-structure} to define a termination condition based on the achieved training cost, if a sufficiently large training data set is used.

\subsubsection{Local cost function}

While tensor products of Haar-random single-qubit states are relatively cheap to prepare, a training cost of the form used in Theorem~\ref{thm:product-training-global-cost-appendix} and Corollary~\ref{corol:product-training-global-cost-variable-structure} also can be challenging to deal with. 
Namely, it is based on performing measurements on all $n$ qubits at the output of a parametrized quantum circuits, which leads to poor trainability due to barren plateaus even for QNNs of moderate depth~\cite{cerezo2020cost}. This occurs because the loss observables $O^{\mathrm{loss}}_{\ket{\Psi_{\rm P}^{(j)}}, \ket{\Phi_{\rm P}^{(j)}}} = O^{G}_{\ket{\Psi_{\rm P}^{(j)}}, \ket{\Phi_{\rm P}^{(j)}}} = \mathds{1} - \ket{\Psi_{\rm P}^{(j)}}\bra{\Psi_{\rm P}^{(j)}}$ are global.
In contrast, using a cost function that depends only on measurements of single qubits at the circuit output can lead to improved trainability for shallow QNNs~\cite{cerezo2020cost}. 

Therefore, we consider a local version of our cost based on Haar-random single-qubit states. We still employ the product state training data 
\begin{align}
    \mathcal{D}_{\rm P}(N)
    = \{(\ket{\Psi_{\rm P}^{(j)}},\ket{\Phi^{(j)}_{\rm P}}) \}_{j=1}^{N}
    = \left\{\left(\bigotimes_{i=1}^n \ket{\psi^{(j)}_i}, U\left(\bigotimes_{i=1}^n \ket{\psi^{(j)}_i}\right)\right)\right\}_{j=1}^N \, ,
\end{align}
with the $\ket{\psi^{(j)}_i}$ drawn i.i.d.~according to ${\rm Haar}_1$. Now, we define the local loss observables
\begin{align}
    O^{\mathrm{loss}}_{\ket{\Psi_{\rm P}^{(j)}}, \ket{\Phi_{\rm P}^{(j)}}}
    &= O^{L}_{\ket{\Psi_{\rm P}^{(j)}},\ket{\Phi_{\rm P}^{(j)}}}
    = \mathds{1} - \frac{1}{n}\sum\limits_{i=1}^n \ket{\psi_i^{(j)}} \bra{\psi_i^{(j)} }\otimes \mathds{1}_{\bar{i}} \, ,
\end{align}
which again have operator norm bounded by $1$. 
Here $\mathds{1}_{\bar{i}}$ denotes the identity acting on all but the $i$th qubit. With this choice of training data and loss observables, Eq.~\eqref{eq:unitary-compiling-cost-general} gives the training cost
\begin{align}
    C_{ \mathcal{D}_{\rm P}(N)}^{L}(\alv)
    &= \frac{1}{N}\sum\limits_{j=1}^N \Tr\left[V_{\vec{\alpha}}^\dagger\ket{\Phi^{(j)}_{\rm P}}\bra{\Phi^{(j)}_{\rm P}}V_{\vec{\alpha}}\cdot O^{L}_{\ket{\Psi_{\rm P}}, \ket{\Phi_{\rm P}}}\right]\\
    &= 1 -\frac{1}{nN}\sum\limits_{j=1}^N\sum\limits_{i=1}^n \Tr \left[ U \ket{\Psi_{\rm P}^{(j)}}\bra{\Psi_{\rm P}^{(j)}}  U^\dagger V_{\Delta t} ( \alv )\left(\ket{\psi_i^{(j)}} \bra{\psi_i^{(j)} }\otimes \mathds{1}_{\bar{i}}\right) V_{\Delta t} ( \alv )^\dagger\right] \, .
\end{align}
Comparing this expression to $C_{ \mathcal{D}_{\rm P}(N)}^{G}(\alv)$, we see that $C_{ \mathcal{D}_{\rm P}(N)}^{L}(\alv)$ still uses tensor products of random single-qubit states as inputs, but instead of measuring all qubits at the output, we average over simple single-qubit measurements.
And the expected cost arising from Eq.~\eqref{eq:unitary-compiling-expected-cost-general} is
\begin{align}
    C_{{\rm Haar}_1^{\otimes n}}^{L}(\alv) 
    &= \mathbb{E}_{\ket{\Psi_{\rm P}} \sim {\rm Haar}_1^{\otimes n}} \left[\Tr\left[ V_{\Delta t} ( \alv )^\dagger\ket{\Phi_{\rm P}}\bra{\Phi_{\rm P}}V_{\Delta t} ( \alv )\cdot O^{L}_{\ket{\Psi_{\rm P}},\ket{\Phi_{\rm P}}}\right] \right] \label{eq:product-local-cost}\\
    &= 1 - \mathbb{E}_{\ket{\Psi_{\rm P}}\sim {\rm Haar}_1^{\otimes n}}\left[ \frac{1}{n}\sum\limits_{i=1}^n  \Tr \left[ U \ket{\Psi_{\rm P}}\bra{\Psi_{\rm P}}  U^\dagger V_{\Delta t} ( \alv ) \left( \ket{\psi_i^{(j)}} \bra{\psi_i^{(j)} }\otimes \mathds{1}_{\bar{i}} \right)  V_{\Delta t} ( \alv )^\dagger\right]\right] \, .
\end{align}

The first observation underlying our analysis of the overall simulation fidelity in terms of $C_{{\rm Haar}_1^{\otimes n}}^{L}(\alv)$ is that the global and local versions of the cost defined in terms of tensor products of Haar-random single-qubit states are tightly related:

\begin{lem}[Product cost: Global versus local]\label{lem:product-cost-global-versus-local}
    For any parameter setting $\alv$, we have the following relation between the global and local versions of the product cost on $n$ qubits:
    \begin{equation}
        C_{{\rm Haar}_1^{\otimes n}}^{L}(\alv)
        \leq C_{{\rm Haar}_1^{\otimes n}}^{G}(\alv)
        \leq n\cdot C_{{\rm Haar}_1^{\otimes n}}^{L}(\alv) \, .
    \end{equation}
\end{lem}

\begin{proof}
The proof of this Lemma is a direct application of the reasoning from Appendix C of~\cite{khatri2019quantum}. 
\end{proof}

This relation between the two costs allows us to extend our previous analysis to $C_{{\rm Haar}_1^{\otimes n}}^{L}(\alv)$:

\begin{corol}[Simulation error (product training, local cost) -- Second half of Theorem~\ref{thm:product-training-global-cost}]\label{corol:product-training-local-cost-appendix}
    For a QNN $V_{\Delta t} ( \alv )$ with $K$ parameterized local gates, with probability $\geq 1-\delta$ over the choice of training data $\mathcal{D}_{\rm P}(N)=\left\{ \bigotimes_{j=1}^n \ket{\psi_i^{(j)}}, U \bigotimes_{j=1}^n \ket{\psi_i^{(j)}} \right\}_{j=1}^N$, with the $\ket{\psi_i^{(j)}}$ independent Haar-random single-qubit states, the total simulation fidelity satisfies
    \begin{align}
        \overline{F}(\alv_{\rm opt}, M\cdot\Delta t)
        &\geq 1 - 2M^2\left[\frac{\epsilon^2}{d+1} + 4n C_{\mathcal{D}_{\rm P}(N)}^{L}(\alv_{\rm opt}) \right]
        - \mathcal{O} \left(M^2n\left(\sqrt{\frac{K\log\left(K\right)}{N}}+ \sqrt{\frac{\log(\nicefrac{1}{\delta})}{N}} \right)\right) \, ,
    \end{align}
    where $\alv_{\rm opt} = \alv_{\rm opt} (\mathcal{D}_{\rm P}(N))$ denotes the final parameter setting after training.
\end{corol}

\begin{proof}
    We start by stating the generalization guarantee obtained from Theorem~\ref{thm:generalization-bound-general} and Corollary~\ref{corol:generalization-bound-unitary-compilation-general} for our scenario: With probability $\geq 1-\delta$ over the choice of training data $\mathcal{D}_{\rm P}(N)=\left\{ \bigotimes_{i=1}^n \ket{\psi_i^{(j)}}, U \bigotimes_{j=1}^n \ket{\psi_i^{(j)}} \right\}_{i=1}^N$, with the $\ket{\psi_i^{(j)}}$ independent Haar-random single-qubit states, the final parameter setting $\alv_{\rm opt} = \alv_{\rm opt}(\mathcal{D}_{\rm P}(N))$ satisfies
    \begin{align}
        C_{{\rm Haar}_1^{\otimes n}}^{L}(\alv) 
        &\leq C_{\mathcal{D}_{\rm P}(N)}^{L}(\alv_{\rm opt}) + \mathcal{O} \left(\sqrt{\frac{K\log\left(K\right)}{N}} + \sqrt{\frac{\log(\nicefrac{1}{\delta})}{N}} \right)\, .
    \end{align}
    Combining Lemma~\ref{lem:entangled-global-versus-average-fidelity-versus-hst}, Lemma~\ref{lem:product-versus-entangled-cost}, and Lemma~\ref{lem:product-cost-global-versus-local}, this implies that, with probability $\geq 1-\delta$,
    \begin{align}
        C_{\mbox {\tiny \rm HST}}(U, V_{\Delta t} (\alv_{\rm opt}))
        &=\frac{d+1}{d}C_{{\rm Haar}_n}^{G}(\alv)\\
        &\leq 2 C_{{\rm Haar}_1^{\otimes n}}^{G}(\alv)\\
        &\leq 2nC_{{\rm Haar}_1^{\otimes n}}^{L}(\alv)\\
        &\leq 2nC_{\mathcal{D}_{\rm P}(N)}^{L}(\alv_{\rm opt}) + \mathcal{O} \left(n\left(\sqrt{\frac{K\log\left(K\right)}{N}}+ \sqrt{\frac{\log(\nicefrac{1}{\delta})}{N}} \right)\right) \, .
    \end{align}
    We again recall that, by Eq.~\eqref{eq:overall-simulation-fidelity-bound-via-hst-cost},
    \begin{align}
        \overline{F}(\alv_{\rm opt}, M\cdot\Delta t)
        &\geq  1 - \frac{2M^2}{d+1}\left[\epsilon^2 + 2d\cdot C_{\mbox {\tiny \rm HST}}(U, V_{\Delta t} (\alv_{\rm opt}))\right] \, .
    \end{align}
    Plugging our above upper bound on the HST cost $C_{\mbox {\tiny \rm HST}}(U, V_{\Delta t} (\alv_{\rm opt}))$ into this lower bound for the overall simulation fidelity $\overline{F}(\alv_{\rm opt}, M\cdot\Delta t)$, we obtain: With probability $\geq 1-\delta$,
    \begin{align}
        \overline{F}(\alv_{\rm opt}, M\cdot\Delta t)
        &\geq  1 - \frac{2M^2}{d+1}\left[\epsilon^2 + 2d\cdot \left( 2nC_{\mathcal{D}_{\rm P}(N)}^{L}(\alv_{\rm opt})  + \mathcal{O} \left(n\left(\sqrt{\frac{K\log\left(K\right)}{N}}+ \sqrt{\frac{\log(\nicefrac{1}{\delta})}{N}} \right)\right)\right)\right]\\
        &= 1 - 2M^2\left[\frac{\epsilon^2}{d+1} + \frac{4d}{d+1}\cdot n C_{\mathcal{D}_{\rm P}(N)}^{G}(\alv_{\rm opt})\right]
        - \mathcal{O} \left(M^2n\cdot\frac{d}{d+1}\cdot\left(\sqrt{\frac{K\log\left(K\right)}{N}}+ \sqrt{\frac{\log(\nicefrac{1}{\delta})}{N}} \right)\right) \\
        &\geq 1 - 2M^2\left[\frac{\epsilon^2}{d+1} + 4 n C_{\mathcal{D}_{\rm P}(N)}^{G}(\alv_{\rm opt})\right]
        - \mathcal{O} \left(M^2 n\left(\sqrt{\frac{K\log\left(K\right)}{N}}+ \sqrt{\frac{\log(\nicefrac{1}{\delta})}{N}} \right)\right)\, ,
    \end{align}
    which is the claimed bound.
\end{proof}

Again, we can easily state the variable structure version of this Theorem:

\begin{corol}\label{corol:product-training-local-cost-variable-structure}
    For a QNN $V_{\Delta t} ( \alv )$ with a variable structure such that, for every $\kappa\in\mathbb{N}$, there are at most $G_\kappa\in\mathbb{N}$ allowed structures with exactly $\kappa$ parameterized local gates, with probability $\geq 1-\delta$ over the choice of training data $\mathcal{D}_{\rm P}(N)=\left\{ \bigotimes_{j=1}^n \ket{\psi_i^{(j)}}, U \bigotimes_{j=1}^n \ket{\psi_i^{(j)}} \right\}_{i=1}^N$, with the $\ket{\psi_i^{(j)}}$ independent Haar-random single-qubit states, the total simulation fidelity satisfies
    \begin{align}
        \overline{F}(\alv_{\rm opt}, M\cdot\Delta t)
        &\geq 1 - 2M^2\left[\frac{\epsilon^2}{d+1} + 4 n C_{\mathcal{D}_{\rm P}(N)}^{L}(\alv_{\rm opt})\right]
        - \mathcal{O} \left(M^2 n \left(\sqrt{\frac{K\log\left(K\right)}{N}}+ \sqrt{\frac{\log G_K}{N}} + \sqrt{\frac{\log(\nicefrac{1}{\delta})}{N}} \right)\right) \, ,
    \end{align}
    where $K=K(\mathcal{D}_{\rm P}(N))$ denotes the number of parameterized local gates used in the final structure after training and $\alv_{\rm opt} = \alv_{\rm opt} (\mathcal{D}_{\rm P}(N))$ denotes the final parameter setting after training.
\end{corol}

As before, from  Corollaries~\ref{corol:product-training-local-cost-appendix} and~\ref{corol:product-training-local-cost-variable-structure} we can derive termination conditions based on the training cost achieved in the optimization and the size of the training data sets.

\begin{rmk}
    As the results of Ref.~\cite{caro2022outofdistribution} also apply when replacing the data-generating measure ${\rm Haar}_1^n$ by a tensor product of $1$-qubit $2$-designs, we can do the same replacement here and still obtain the same bounds.
    As a concrete example, the same guarantees hold when training on tensor products of uniformly random single-qubit stabilizer states.
\end{rmk}

\section{Further details on numerical implementations}\label{Ap:Numerics}

\subsection{Statement of Algorithm}

\title{Resource Efficient Fast-Forwarding algorithm (Analytic)}

\SetKwInput{KwInput}{Input}                
\SetKwInput{KwOutput}{Output}              

\begin{algorithm}
\DontPrintSemicolon
  
  \KwInput{$n$-qubit Hamiltonian, $H$; Target simulation fidelity at $M_0$ fast-forwarding steps, $1- \varepsilon$; Diagonal ansatz, $V_t(\thv, \gamv) = W(\thv) D(t\gamv) W(\thv)^{\dagger}$ with $K$ parameterized gates; set of $N\sim M_0^4 \tfrac{K\log (K)}{(\varepsilon/2)^2}$ input-output training states, $\mathcal{D}(N) = \{(\ket{\Psi^{(j)}},\ket{\Phi^{(j)}}) \}_{j=1}^{N}$ where the input states are random product states $\{\bigotimes_{i=1}^n \ket{\psi_i^{(j)}}\}_{j=1}^N$, with $\ket{\psi_i^{(j)}}$ drawn from the 1-qubit Haar distribution, and the output states are generated by evolving the input states by a short time Trotterized evolution operator $\ket{\Phi^{(j)}}=U_{\Delta t}\ket{\Psi^{(j)}}$ where $U_{\Delta t}\approx e^{-i H \Delta t}$. }
  \KwOutput{Time-Dependent QNN, $V_t(\theta_{\rm opt}, \gamma_{\rm opt})$}
  
  Randomly initialize the parameters of $V_t(\thv, \gamv)$ \\
  \While {$C_{\mathcal{D}_{\rm P}(N)}^{G}(\thv, \gamv) > \tfrac{\varepsilon}{16 M_0^2} - \tfrac{\epsilon^2}{4(2^n+1)}$}
    {Calculate gradient vector using Equations \eqref{eq:CostEigenvectorGradient} \& \eqref{eq:CostDiagonalGradient} via the circuit shown in Fig.~\ref{fig:REFFCircuit}.  \\ Update parameters of $V_{\Delta t}(\thv, \gamv)$ with classical optimizer \\
    }
    \Return $(\theta_{\rm opt}, \gamma_{\rm opt})$

\caption{Resource Efficient Fast-Forwarding}
\label{alg:REFF}
\end{algorithm}

The REFF algorithm is outlined with pseudocode in Alg.~\ref{alg:REFF}. By inverting Theorem \ref{thm:product-training-global-cost}, our results provide an upper bound on the number of training states required to achieve a target simulation fidelity, $1-\varepsilon$, after $M_0$ fast-forwarding steps at $N\sim M_0^4 \tfrac{K\log (K)}{(\varepsilon/2)^2}$. Our numerical results, however, have found that a full generalization can be achieved with far fewer states than this bound would suggest. To take advantage of this and seek a minimally sized training dataset, an algorithm variant is motivated, starting with a small training dataset that is grown over time until generalization is observed. As we are working below the upper bound, the guarantees on generalization no longer apply, so an extra validation error is required to quantify the QNN's generalization. This could be implemented by creating a validation dataset of Haar-random product states that the $C_{\rm REFF}$ cost is periodically tested against as the optimization progresses. If the training $C_{\rm REFF}$ is decreasing during the optimization, but the validation $C_{\rm REFF}$ is observed to plateau, this indicates the training dataset is too small and its size should be increased.


\begin{figure}
\centering
\includegraphics[width =\columnwidth]{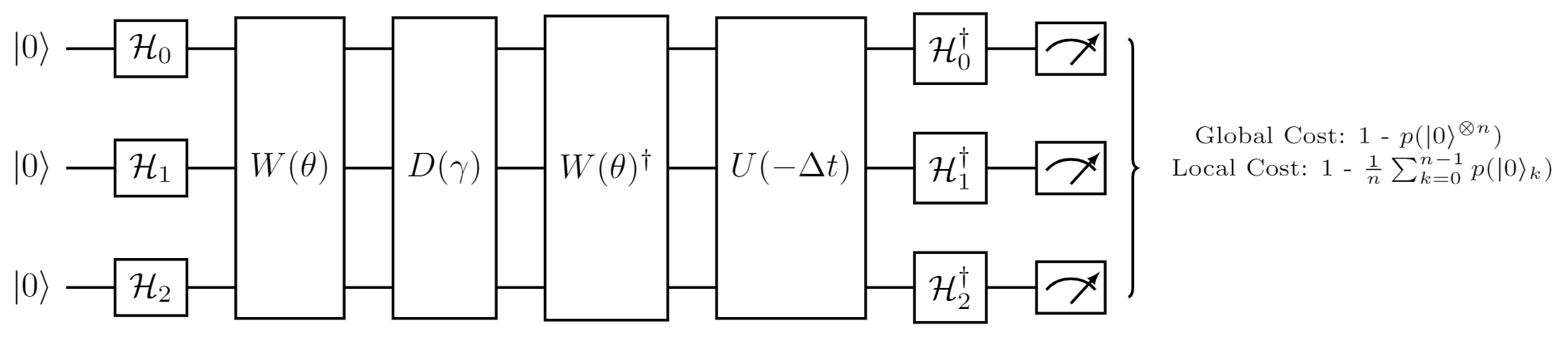}
\vspace{-6mm}
\caption{\small \textbf{REFF Circuit.} A generalized 3-qubit representation of the circuit used to compute $C_{\rm REFF}$ is displayed. $\mathcal{H}_k$ is a Haar-random single-qubit unitary. For a training dataset of size $N$, the output of this circuit is averaged over the $N$ unique training states to compute the cost function.}
\label{fig:REFFCircuit}
\end{figure}

\subsection{Gradient Formula}

Following the method of \cite{cirstoiu2020variational, gibbs2021long}, the partial derivative of $C_{\rm REFF}(U,V(\thv, \gamv))$ with respect to $\theta_l$, is
\begin{equation} \label{eq:CostEigenvectorGradient}
    \begin{split}
        \frac{\partial C_{\rm REFF}(U,V)}{\partial \theta_l} = \frac{1}{2} \Big( & C_{\rm REFF}(U, W_{l+} D W^\dagger) \,
        - \, C_{\rm REFF}(U, W_{l-} D W^\dagger) \\
        + \ & C_{\rm REFF}(U, W D (W_{l+})^\dagger) \
        - \  C_{\rm REFF}(U, W D (W_{l-})^{\dagger}) \Big) \ .
    \end{split}
\end{equation}

The unitary $W_{l+}$ ($W_{l-}$) is generated from the original unitary $W(\thv)$ by the addition of an extra $\frac{\pi}{2}$ ($-\frac{\pi}{2}$) rotation about a given parameter's rotation axis:
\begin{equation}
    W_{l\pm} := W\left( \thv_{l\pm} \right) \ \ \text{with}  \ \ (\theta_{l\pm})_i := \theta_l \pm \frac{\pi}{2} \delta_{i,l} \; .
\end{equation}

The analogous formula for the partial derivative with respect to $\gamma_l$, is
\begin{equation} \label{eq:CostDiagonalGradient}
    \begin{split}
        \frac{\partial C_{\rm REFF}(U,V)}{\partial \gamma_l} = \frac{1}{2} \Big( & C_{\rm REFF}(U, W D_{l+} W^\dagger) \,
        - \, C_{\rm REFF}(U, W D_{l-} W^\dagger) \Big) \ .
    \end{split}
\end{equation}

\subsection{Choice of Ansatz}

\paragraph*{XY Hamiltonian.}

For our diagonalizing unitary, we exploit the particle-number conservation property of the Heisenberg Hamiltonians, and for $W$ use only 2-qubit Givens rotation gates that also respect this symmetry. This ensures the state remains in the symmetry subspace, simplifying the optimization. For the numerics shown in Fig.\ref{fig:main_numerics}b), a brickwork-style ansatz was used for $W$. Each layer was composed of a Givens rotation gate between odd pairs of qubits, then a Givens rotation gate between even pairs of qubits, resulting in $n-1$ gates per layer. All the examples used $1.5n$ layers, resulting in a total gate count for $W$ of $1.5n(n-1)$ and a depth of $3n$; this can be considered an improvement to the $\mathcal{O}\left(n^2\right)$ gate count and $\mathcal{O}\left(n\log\left(n\right)\right)$ circuit depth presented in \cite{verstraete2009quantum}. The ansatz for $D$ was composed of a single $R_z$ gate on each qubit.

\paragraph*{Heisenberg Hamiltonian.}
It is known \cite{anselmetti2021local} that a 2-local nearest-neighbor gate fabric of Givens rotations is not universal for the Hamming-weight preserving subgroup $\mathcal{H}(2^n)$, so the ansatz we used for the XY Hamiltonian does not diagonalize the more general Heisenberg Hamiltonian. In these numerics, for the $W$ layered ansatz, we instead use the most general 2-qubit gate that conserves the particle number, which takes 4 parameters and is displayed below.

\begin{equation}
Sym(\theta_1, \theta_2, \theta_3, \theta_4) = 
\begin{pmatrix}
    e^{i \theta_1} & 0 & 0 & 0  \\
    0 & \cos(\theta_2) & -e^{i \theta_3} \sin(\theta_2) & 0 \\
    0 & e^{i \theta_4} \sin(\theta_2) & e^{i (\theta_3 + \theta_4)} \cos(\theta_2) & 0 \\
    0 & 0 & 0 & 1
\end{pmatrix}
\end{equation}

$D$ is composed of an $R_z$ rotation gate on each qubit, and an $R_{zz}$ gate between all pairs of qubits.

\medskip

\begin{figure}
\centering
\includegraphics[width =0.55\columnwidth]{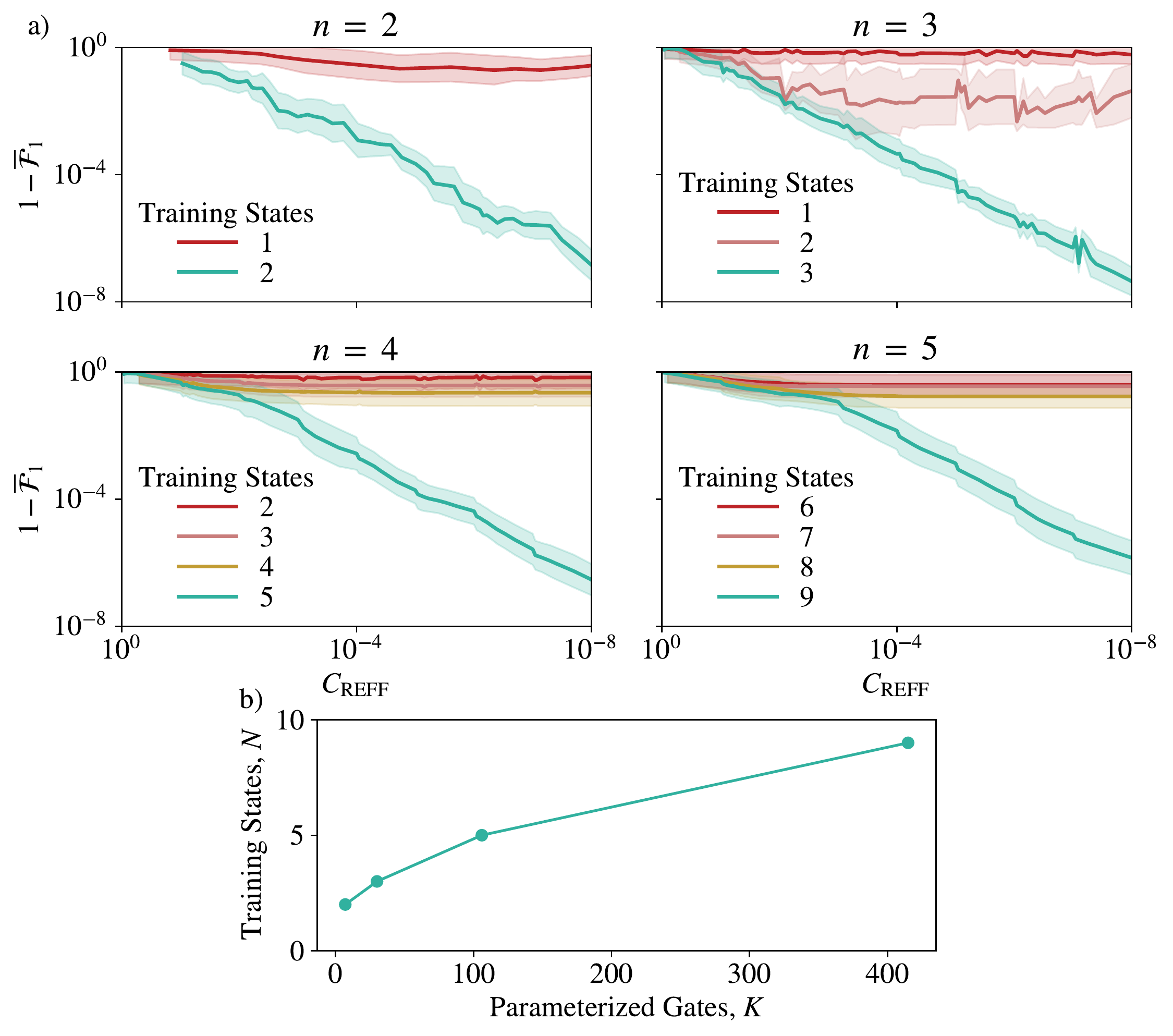}
\vspace{-3mm}
\caption{\small \textbf{Scaling.} a) REFF is applied to increasing sizes of the $n$-qubit Heisenberg Hamiltonian, $H = \sum_{i = 1}^n \vec{S}_i \cdot \vec{S}_{i+1}$ with periodic boundary conditions. An increasing size dataset of Haar-random product states is used for training, to determine the minimum number required for generalization. This is seen when minimizing the training loss $(C^{\rm G}_{\rm REFF})$ results in simultaneously minimizing the validation loss ($1 - \overline{\mathcal{F}}_1$) without premature plateauing. b) The minimum number of training states required for generalization in the above numerics are plotted against the number of parameterized gates in its VFF ansatz, showing an apparent sub-linear scaling.
}
\label{fig:ParameterScaling}
\end{figure}

\subsection{Data Requirements for Generalization}

Theorem \ref{thm:product-training-global-cost} upper-bounds the number of training states required to guarantee a high simulation fidelity for a QNN with K parameterized gates by the scaling $\mathcal{O}(K\log(K))$. Here we numerically investigate this bound by performing REFF on increasing $n$-qubit Heisenberg Hamiltonians and observe the minimum number of training states required for generalization. For each system size, we perform REFF with a range of training set sizes, training until $C^{\rm G}_{\rm REFF} = 10^{-8}$ and simultaneously compute the simulation fidelity ($1 - \overline{\mathcal{F}}_1$) as our metric for generalization. Each instance of REFF performed on a particular training set size was repeated 5 times to generate a mean and standard deviation. We observe, as shown in Fig.~\ref{fig:ParameterScaling}, that training sets of insufficient size do not produce generalization when trained on, but when the training set reaches some critical threshold, minimizing $C^{\rm G}_{\rm REFF}$ simultaneously minimizes the simulation fidelity. The scaling of this threshold is plotted in Fig.~\ref{fig:ParameterScaling}b); as a function of the number of parameterized gates, $K$, the required number of training states, $N$, appears to scale sub-linearly, within our analytic upper-bound.

\subsection{Simulation fidelity against the true Hamiltonian evolution}

The numerical results in Fig. \ref{fig:main_numerics} demonstrate that with the correct ansatz design, the Hilbert-Schmidt Test between the Trotterized unitary and the diagonalized ansatz can reduced arbitrarily small for the systems tested. As shown in the inset of Fig. \ref{fig:main_numerics}b), this allows the action of the Trotterized unitary to be fast-forwarded for long-time simulations. The unitary we are learning is a Trotterization, with an associated Trotter error with respect to the true Hamiltonian evolution, which will also be learned during the REFF training. REFF can at best perfectly learn the evolution with this Trotter error included, however increasing the order of the Trotter approximation decreases this error to reproduce the true Schr\"{o}dinger evolution.

\begin{figure}
\centering
\includegraphics[width =0.55\columnwidth]{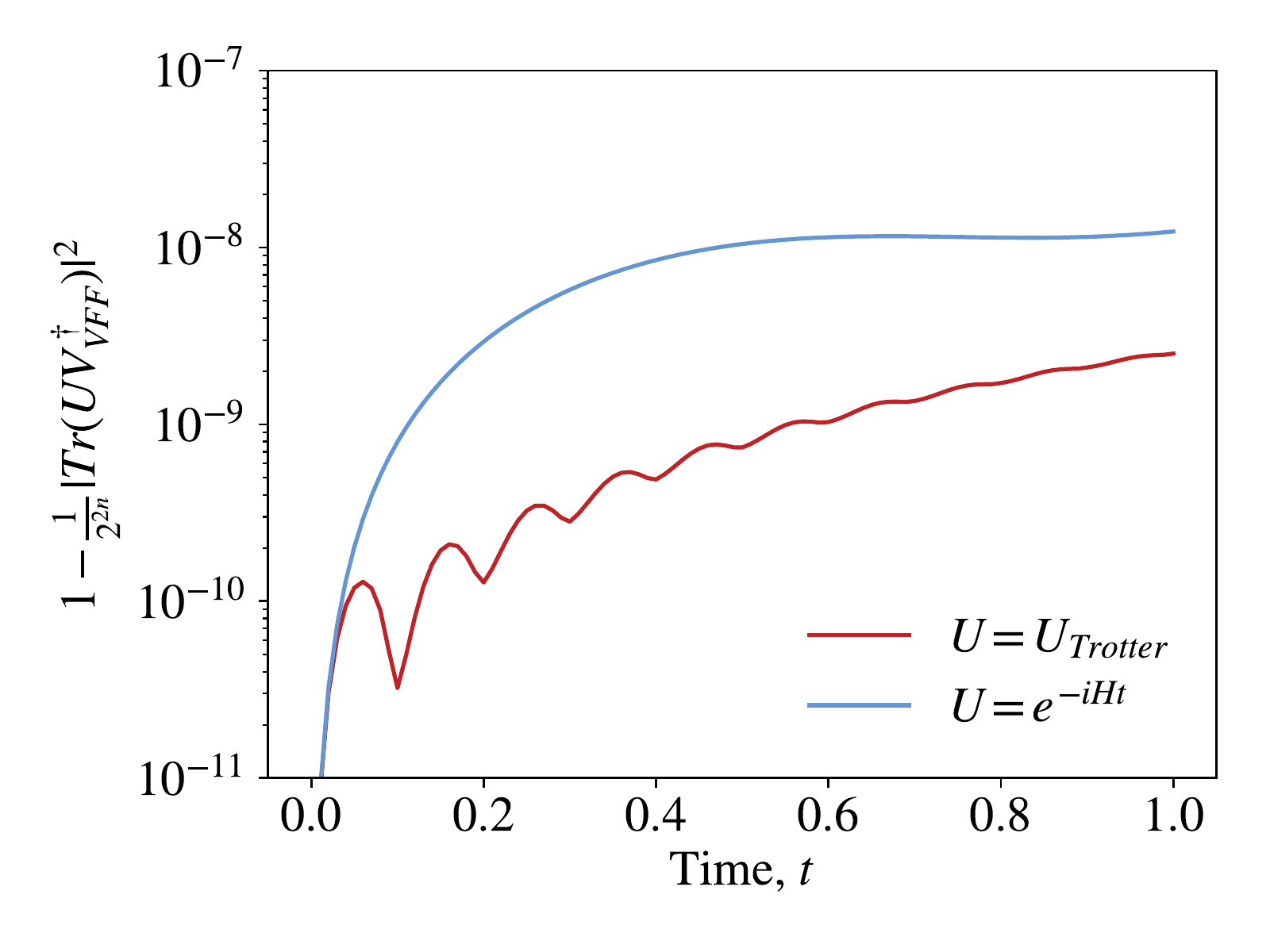}
\vspace{-3mm}
\caption{\small \textbf{Trotter Error.} The REFF algorithm is applied to the 10 qubit Hamiltonian $H=\sum_{i=1}^{9} X_i X_{i+1} + Y_i Y_{i+1}$ with open boundary conditions, and the resulting diagonalized unitary performs a fast-forwarding of the Hamiltonian. The Trotterization has a timestep of $\Delta t = 0.1$. The fidelity of the fast-forwarding is computed with respect to the Trotterized unitary and the true Hamiltonian evolution.} 
\label{fig:TrueEvolution}
\end{figure}

Fig. \ref{fig:TrueEvolution} shows the fast-forwarding of a diagonalized ansatz after REFF training against the 10 qubit Hamiltonian $H=\sum_{i=1}^{9} X_i X_{i+1} + Y_i Y_{i+1}$ with open boundary conditions. To compare the simulation fidelity against the Trotter unitary $U(\Delta t)$ for non-integer multiples of $\Delta t$, for $t = M\Delta t + c$ (with $c < \Delta t$) the unitary applied is $U_{\text{Trotter}}(t) = U(\Delta t)^M U(\frac{c}{\Delta t})$.
Here the Trotterization is a 2nd order Suzuki-Trotter approximation, with Trotter number 10, and hence has a very low Trotter error. The VFF ansatz had in total 360 CNOT gates, compared to 380 in the Trotter unitary. 

\subsection{Quantum state tomography of hardware implementation}
Fig.~\ref{fig:HardwareFastForwarding}b) shows the evaluation of the fast-forwarding performance of the QNN trained on ibmq\_bogota. We use quantum state tomography to reconstruct the output density matrix of both the REFF-evolved state and the Trotter-evolved state at timestep $N$, $\rho(N)$, and compute the fidelity with respect to the noise-free Trotter evolved state, $|\psi(N)\rangle$. The method for performing this follows \cite{gibbs2021long}; an $n$-dimensional density matrix $\rho$ can be decomposed into the Pauli product basis as $\rho = \boldsymbol \eta \cdot \boldsymbol \sigma^{(n)}$. Here $\boldsymbol \sigma^{(n)}$ is a $4^n$ dimensional vector composed of the elements of the $n$-qubit Pauli group $P_n = \{\sigma_I, \sigma_X, \sigma_Y, \sigma_Z \}^{\otimes n} $ and $\boldsymbol  \eta$ is the corresponding vector of Pauli weights, i.e. $\eta_k = \frac{1}{2^n}\text{Tr}(\sigma^{(n)}_k \rho)$. The values $\eta_k$ were computed on the quantum device using 8192 shots, then used to classically calculate the fidelity $F(N) = \langle\psi(N)|\rho(N)|\psi(N)\rangle$ .

\clearpage

\end{document}